\documentclass[11pt, reprint,prd,superscriptaddress,nofootinbib,floatfix,preprintnumbers]{revtex4}
\usepackage{graphicx, bm, color, amssymb, amsmath,subfigure}

\newcommand{\CP}{C\!P}
\newcommand{\maf}[1]{\mathbf{#1}}

\begin{document}
\preprint{MAN/HEP/2014/11}
\preprint{
UMD-PP-014-014}
\title{Leptogenesis Constraints on the Mass of Right-handed Gauge Bosons}

\author{P. S. Bhupal Dev}
\affiliation{Consortium for Fundamental Physics, School of Physics and Astronomy, University of Manchester, Manchester M13 9PL, United Kingdom}

\author{Chang-Hun Lee}
\affiliation{Maryland Center for Fundamental Physics and Department of Physics, University of Maryland, College Park, Maryland 20742, USA}

\author{R. N. Mohapatra}
\affiliation{Maryland Center for Fundamental Physics and Department of Physics, University of Maryland, College Park, Maryland 20742, USA}

\begin{abstract}
We discuss leptogenesis constraints on the mass of the right-handed $W$-boson ($W_R$)  in a TeV-scale Left-Right seesaw model (LRSM) for neutrino masses. For generic Dirac mass of the neutrinos, i.e.~with all Yukawa couplings $\lesssim 10^{-5.5}$, it has been pointed out that successful leptogenesis requires a lower bound of 18 TeV on the $W_R$ mass, pushing it beyond the reach of LHC. Such TeV-scale LRSM must, however, be parity-asymmetric for type-I seesaw to give the observed neutrino masses. This class of models can accommodate larger Yukawa couplings, which give simultaneous fits to charged-lepton and neutrino masses, by invoking either  cancellations or specific symmetry-textures for Dirac ($M_D$) and Majorana ($M_N$) masses in the seesaw formula. We show that in this case, the leptogenesis bound on $M_{W_R}$ can be substantially weaker,~i.e. $M_{W_R}\gtrsim 3$ TeV  for $M_N \lesssim M_{W_R}$. This happens due to considerable reduction of the dilution effects from $W_R$-mediated decays and scatterings, while the washout effects
due to inverse decays are under control for certain parameter ranges of the Yukawa couplings.
We also show that this model is consistent with all other low energy constraints, such as lepton flavor violation and neutrinoless double beta decay. Thus, a discovery of the right-handed gauge bosons alone at the LHC will not falsify leptogenesis as the mechanism behind the matter-antimatter asymmetry in our Universe.
\end{abstract}

\maketitle
\section{Introduction}\label{sec:1}
Seesaw mechanism~\cite{seesaw} provides a very simple way to understand the smallness of neutrino masses. Two main ingredients of this mechanism are: (a) the introduction of
right-handed (RH) neutrinos $N$ to the Standard Model (SM), and (b) endowing the $N$'s with a Majorana mass which breaks the accidental global $B-L$ symmetry of the SM. In the context of the SM electroweak gauge group $SU(2)_L\times U(1)_Y$, these two features do not follow from any underlying principle, but are rather put in by hand. There are two classes of ultraviolet-complete theories where both these ingredients of seesaw emerge in a natural manner: (i)  the Left-Right (L-R) symmetric theories of weak interactions based on the gauge group $SU(2)_L\times SU(2)_R\times U(1)_{B-L}$~\cite{LR}, and (ii) $SO(10)$ Grand Unified Theory for all interactions~\cite{so10}. The existence of the RH neutrinos is guaranteed by the gauge symmetry and anomaly cancellation in both cases, and their Majorana masses are connected to the breaking scale of local $B-L$ symmetry, which is a subgroup of the above gauge groups. Furthermore, they also predict the number of $N$'s to be three, corresponding to three generations of fermions in the SM. Thus,  the essential ingredients of seesaw are no more adhoc, but are rather connected to symmetries of the extended theory. It is then important to explore how new features of these symmetries can be probed in laboratory experiments. Our focus in this article is on low-scale L-R symmetric models (LRSM), where the seesaw scale can be in the few TeV range, accessible to the Large Hadron Collider (LHC), while satisfying the observed charged-lepton and neutrino mass spectra.

The first question for such models is how the small neutrino masses can be understood if the seesaw scale is indeed in the TeV range, since by naive expectations, the Dirac masses are expected to be similar to the  charged-lepton masses, which after seesaw would give rise to too large tau neutrino mass. In the context of the minimal LRSM, this question becomes specially important since the Higgs sector relates the neutrino Yukawa couplings with charged-lepton ones. There are three ways to fit both charged-lepton  and neutrino masses in such TeV-scale LRSM: (i) by choosing one set of the Yukawa couplings to be $\lesssim 10^{-5.5}$ for a particular vacuum expectation value (VEV) assignment for the SM-doublet Higgs fields; (ii) by choosing larger Yukawa couplings and invoking cancellations between Yukawa couplings in the Dirac neutrino mass matrix to get smaller Dirac masses for neutrinos to get seesaw to work, and (iii) by choosing particular symmetry-motivated textures for the Yukawa couplings that guarantee the leading order seesaw contribution to neutrino masses to vanish (an example of such an LRSM is given in~\cite{ours}). We call these models class I, II, and III models respectively.

An attractive feature of the seesaw mechanism is that the same Yukawa couplings that give rise to light neutrino masses can also resolve one of the outstanding puzzles of cosmology, namely, the origin of matter-antimatter asymmetry, via leptogenesis~\cite{lepto} (for reviews, see e.g.~\cite{reviews}).
The key drivers of leptogenesis are the out-of-equilibrium decays of the RH Majorana neutrinos via the modes $N\to L_l \phi$ and $N\to L_l^c\phi^c$, where $L_l=(\nu_l \quad l)_L^{\sf T}$ (with $l=e,\mu,\tau$) are the $SU(2)_L$ lepton doublets, $\phi$ are the Higgs doublets, and the superscript $c$ denotes the $\CP$-conjugate.  In the presence of $\CP$ violation in the Yukawa sector, these decays can lead to a dynamical lepton asymmetry in the early Universe, satisfying the three Sakharov conditions~\cite{sakharov}. This asymmetry will undergo thermodynamic evolution as the Universe expands and different reactions present in the model have their impact on washing out part of the asymmetry. The remaining final lepton asymmetry is converted to the baryon asymmetry via sphaleron processes~\cite{Kuzmin:1985mm} before the electroweak phase transition. There is also a weak connection between the $\CP$ violation in neutrino oscillations and the amount of lepton asymmetry generated~\cite{reviews}.

 For TeV-scale seesaw models, the generation of adequate lepton asymmetry requires one to invoke resonant leptogenesis~\cite{Pilaftsis:1997dr, Pilaftsis:2003gt}, where
at least two of the heavy  neutrinos have a  small mass difference
comparable to their  decay  widths. In this case, the   heavy  Majorana neutrino   self-energy
contributions~\cite{Liu:1993tg}  to  the  leptonic $\CP$-asymmetry  become
dominant~\cite{Flanz:1994yx} and get resonantly enhanced,
even up to order one~\cite{Pilaftsis:1997dr}.
In the context of an embedding of seesaw into TeV-scale LRSM, there are additional complications due to the presence of RH gauge interactions~\cite{Carlier:1999ac} that contribute to the dilution and washout of the primordial lepton asymmetry generated via resonant leptogenesis.
This was explored in detail in~\cite{hambye}, where it was pointed out that
there is significant dilution of the primordial lepton asymmetry due to $\Delta L=1$ scattering processes such as $Ne_R\leftrightarrow q\bar{q}'$ mediated by $W_R$. This leads to an extra suppression of the final lepton symmetry, in addition to the usual inverse decay $L\phi \to N$ and $\Delta L=2~(0)$ scattering processes $L\phi \leftrightarrow L^c\phi^c~(L\phi)$ present in generic SM seesaw scenarios. This additional dilution factor (also sometimes called efficiency) in this case
is $\kappa\sim \frac{Y^2 M^4_{W_R}}{g_R^4 M^4_N}$, which for $M_N\sim {\rm TeV}$ and $M_{W_R}\sim 3$~TeV can be easily  $\lesssim 10^{-9}$ for generic Yukawa couplings $Y\simeq 10^{-5.5}$. Combined with the dilution effects from inverse decays and entropy, this implies that even for maximal $\CP$ asymmetry $\varepsilon \sim {\cal O}(1)$, the observed baryon to photon ratio $\eta^B_{\rm obs}\sim 6\times 10^{-10}$~\cite{pdg} can be obtained only if  $M_{W_R}\geq 18$ TeV.
 This result is very important because, as argued in~\cite{hambye}, this can provide a way to falsify leptogenesis if a $W_R$ with mass below this limit is observed in experiments.

Since this appears to be a ``make or break'' issue for the idea of leptogenesis, we investigate whether there are {\it any} allowed parameter space in the TeV-scale LRSM where leptogenesis can work with a weakened lower bound on $M_{W_R}$, without being in conflict with observed neutrino data and charged-lepton masses. We work in a version of the model that is parity-asymmetric at the TeV scale, which is anyway necessary if type-I seesaw is responsible for neutrino masses. According to our classification above, the work of~\cite{hambye} falls into the class I models. We explore whether the lower bound can be weakened in the other classes of models discussed above. It could very well be that if other observations push the Yukawa parameters to the range of class I models, the bound of~\cite{hambye} cannot be avoided, thereby providing a way to disprove leptogenesis at the LHC. However, to see how robust and widely applicable the bound of~\cite{hambye} is, we consider in this paper an example of a model which belongs to class II, i.e. neutrino fits are done by cancellation leading to a specific texture for Dirac masses. Similar analysis can also be performed for class III models~\cite{ours}.

We implement the class II strategy for small neutrino masses in the minimal LRSM with a single bi-doublet field in the lepton sector, where some of the leptonic Yukawa couplings are significantly larger than the canonical value of ${\cal O}(10^{-5.5})$ and the $W_R$ mass is in the few TeV range. As noted above, to get small neutrino masses via type-I seesaw, we invoke cancellation between two Yukawa couplings to generate extra suppression and a particular resulting texture for the Dirac masses. 
We find that due to enhanced Yukawa couplings, the reaction rates of $W_R$-mediated scatterings as well $3$-body decays of RH neutrinos $N\to \ell_Rq\bar{q}'$ become sub-dominant to the rates of
$\CP$-violating 2-body decay modes $N\to L\phi, L^c\phi^c$. As a result, the lower limit on $W_R$ mass can be brought within the LHC reach for a range of Yukawa couplings for which the washout effect due to inverse decays $L\phi\to N$ is in control. New aspects in our work that goes beyond that of~\cite{hambye} are the following: (i) we give a realistic fit for all lepton masses and mixing with larger Yukawa couplings ($\sim 10^{-2}$); (ii) we calculate the primordial $\CP$ asymmetry $\varepsilon$ in our model using the Yukawa couplings demanded by our specific neutrino fit (see text for precise numbers), instead of just assuming the maximal $\CP$-asymmetry $\varepsilon\sim 1$
as done in~\cite{hambye}; and (iii) we take the heavy-neutrino and charged-lepton
flavor effects into account in the thermodynamic evolution of the lepton asymmetry.\footnote{For the importance of flavor effects in resonant leptogenesis, see e.g.~\cite{DMPT} and references therein.} It is a consequence of  (i) and (iii), which leads us to a weaker $W_R$ mass bound from leptogenesis.

This paper is organized as follows: in Section~\ref{sec:2}, we discuss the basic ingredients of
our new TeV-scale LRSM with large Yukawa couplings; in Section~\ref{sec:2a} we present a numerical fit for lepton masses and mixing, and in Section~\ref{sec:2b} we list the model
predictions for low energy observables.
In Section~\ref{sec:3}, we discuss the thermodynamic evolution of the lepton asymmetry, and
present some numerical results based on the fit given in Section~\ref{sec:2a}. We also derive the
new lower bound on $M_{W_R}$ from leptogenesis, which is the main result of our paper.
In Section~\ref{sec:4}, we discuss some future directions, including possible implications of our result for L-R seesaw phenomenology. Our conclusions are given in~\ref{sec:5}. Appendix~\ref{app:A} summarizes the reduced scattering cross sections used in our solution to the Boltzmann equations.

\section{ Review of the new left-right seesaw model}\label{sec:2}
We start by reviewing the basic structure of our model. As in the usual L-R models~\cite{LR}, the  fermions are assigned to the following irreducible representations of the gauge group $SU(2)_L\times SU(2)_R\times U(1)_{B-L}$: denoting
$Q\equiv (u \quad d)^{\sf T}$ and $\psi\equiv (\nu_l \quad l)^{\sf T}$ as the quark and lepton doublets respectively, $Q_L: ({\bf 2},{\bf 1},1/3)$ and $\psi_L: ({\bf 2},{\bf 1}, -1)$ are assigned to doublets under $SU(2)_L$, while $Q_R: ({\bf 1},{\bf 2}, 1/3)$ and $\psi_R: ({\bf 1}, {\bf 2}, -1)$ are doublets under $SU(2)_R$. Note that the $B-L$ quantum numbers directly follow from the definition of the electric charge: $Q=T_{3L}+T_{3R}+(B-L)/2$~\cite{charge}, where $T_{3L}$ and $T_{3R}$ are the third components of isospin under $SU(2)_L$ and $SU(2)_R$ respectively.
The Higgs sector of the model consists of the following multiplets under $SU(2)_L\times SU(2)_R\times U(1)_{B-L}$:
\begin{eqnarray}
\phi\equiv\left(\begin{array}{cc}\phi^0_1 & \phi^+_2\\\phi^-_1 & \phi^0_2\end{array}\right) \; : ({\bf 2}, {\bf 2}, 0),
\qquad \Delta_R\equiv\left(\begin{array}{cc}\Delta^+_R/\sqrt{2} & \Delta^{++}_R\\\Delta^0_R & -\Delta^+_R/\sqrt{2}\end{array}\right) \; : ({\bf 1}, {\bf 3}, 2) \; .
\label{eq:higgs}
\end{eqnarray}
The gauge symmetry $SU(2)_R\times U(1)_{B-L}$ is broken by the triplet VEV $\langle \Delta^0_R\rangle = v_R$ to the group $U(1)_Y$ of the SM, whereas the VEV of the $\phi$ field $\langle\phi\rangle={\rm diag}(\kappa, \kappa')$ breaks the SM gauge group $SU(2)_L\times U(1)_Y$ to $U(1)_{\rm em}$. 

There are versions of the LRSM where parity and $SU(2)_R$ gauge symmetry scales are decoupled so that the LH counterpart ($\Delta_L$) to the $\Delta_R$ field becomes heavy when the discrete parity symmetry is broken, and disappear from the low energy theory~\cite{CMP}. Our model falls into this class of the so-called ``broken $D$-parity" models. The low energy Lagrangian in this case has invariance under the L-R gauge group but not parity.
An important consequence of this is that the type II contribution to neutrino masses~\cite{type2}  are negligible and we can do neutrino fits only with type I seesaw~\cite{seesaw}.

We also wish to note that, in our model, the heavy scalar triplets $\Delta_L({\bf 3},{\bf 1},2)$ can have masses of order $10^{14}$ GeV or higher, so that the type I seesaw term dominates in the neutrino mass formula. The question then  arises regarding its impact on leptogenesis (for reviews on type-II seesaw leptogenesis, see e.g.~\cite{type2lepto}). Whether type II leptogenesis produces a lepton asymmetry from $\Delta_L$ decays does not affect our result, since above the TeV scale, lepton number violating decays of the RH neutrinos are in thermal equilibrium, and therefore, they will wash out any pre-existing lepton asymmetry. The final lepton asymmetry in our model is dominantly produced by the quasi-degenerate heavy neutrino decays via the resonant enhancement mechanism~\cite{Pilaftsis:1997dr, Pilaftsis:2003gt} at temperatures close to the heavy neutrino mass scale, at which the type-II contributions from $\Delta_L$ are hugely Boltzmann-suppressed and can be safely neglected.

To see how the fermions pick up mass and how seesaw mechanism arises in our L-R model,
we write down the generic Yukawa Lagrangian of the model:
\begin{eqnarray}
{\cal L}_Y \ = \ h^{q}_{ij}\bar{Q}_{L,i}\phi Q_{R,j}+\tilde{h}^{q}_{ij}\bar{Q}_{L,i}\tilde{\phi} Q_{R,j}+
h^{l}_{ij}\bar{L}_i\phi R_j 
+ \tilde{h}^{l}_{ij}\bar{L}_i\tilde{\phi}R_j
+f_{ij} (R_iR_j\Delta_R +L_iL_j\Delta_L)+{\rm H.c.},
\label{eq:yuk}
\end{eqnarray}
where $i,j=1,2,3$ stand for the fermion generations, $\tilde{\phi}=\tau_2\phi^*\tau_2$ ($\tau_2$ being the second Pauli matrix) and $L_i~(R_i)$ are short-hand notations for the lepton doublets $\psi_{L(R),i}$. Note that if the theory obeyed exact L-R symmetry, we would have the property that $h=h^\dagger$ and $\tilde{h}=\tilde{h}^\dagger$. Since we will be working in the $D$-parity broken version of the theory, we will assume the Yukawa couplings to be free of these restrictions.
After electroweak symmetry breaking, the Dirac fermion masses are given by the generic formula $M_f~=~h^f\kappa + \tilde{h}^f\kappa'$ for up-type fermions, while for down-type quarks and
charged-leptons, it is the same formula with $\kappa$ and $\kappa'$ interchanged.  The Yukawa Lagrangian in Eq.~(\ref{eq:yuk}) leads to the Dirac mass matrix  for neutrinos $M_D = h^{l}\kappa + \tilde{h^{l}}\kappa'$ and the Majorana mass matrix for the heavy RH  neutrinos $M_N=fv_R$ which go into the seesaw formula
\begin{eqnarray}
M_\nu~\simeq ~-M_DM^{-1}_N M^\mathsf{T}_D
\label{eq:seesaw}
\end{eqnarray}
for calculating the neutrino masses and the heavy-light neutrino mixing.
For $M_N\sim$~few TeV, we need to invoke cancellations between $h\kappa$ and $\tilde{h}\kappa'$ to get small neutrino masses. We use this strategy to illustrate that there exist parameter domains in the theory where the  lower bound on $W_R$ mass from leptogenesis can be weakened. For the RH neutrino mass matrix, we choose the following texture:
\begin{eqnarray}
M_N \ = \ \left(\begin{array}{ccc}  \delta M& M_1 & 0\\M_1 & 0& 0 \\ 0&0&M_2\end{array}\right) \; ,
\label{eq:texture}
\end{eqnarray}
a motivation for this choice being that, for small $\delta M$, it leads ``naturally'' to quasi-degeneracy between the first and second RH neutrinos, as required for resonant leptogenesis. The third RH neutrino mass is kept as a free parameter in our numerical calculations, and could also be degenerate with the other two RH neutrinos. We note here that the small Majorana mass parameter $\delta M$ can arise as a quantum effect at the one loop level in our model, i.e 
\begin{eqnarray}
\delta M \ \sim \ \frac{M_1}{16\pi^2}\left(h^{l}_{1i}\right)^* h^l_{2i} \ln\left(\frac{\Lambda}{M_1} \right)\; ,
\label{rg}
\end{eqnarray}
where $\Lambda$ is some ultraviolet cut-off scale. This is similar to the radiative leptogenesis models~\cite{RGE, Pilaftsis:2005rv, Deppisch:2010fr} where the small mass splitting given by Eq.~(\ref{rg}) is naturally induced by renormalization group running effects from a higher energy scale $\Lambda$.

In L-R models, the charged-lepton mass and Dirac neutrino mass matrices are related, and
fitting the observed neutrino oscillation data simultaneously with the charged-lepton masses is highly nontrivial. As we show below, the model discussed above can reproduce the observed neutrino masses and mixing as well as the charged-fermion masses. Note that while we have presented here
only one example of such a fit, our subsequent analysis and results are generically applicable to other textures with large Yukawa couplings.

\subsection{Neutrino Mass Fit} \label{sec:2a}
In the class of L-R models under consideration, we can always choose a basis of the LH sector prior to $SU(2)_L\times U(1)_Y$ breaking such that the Dirac mass matrix $M_D$ can be written in an upper-triangular form without affecting the RH neutrino texture. In this basis, the mass matrices
obtained from the Yukawa Lagrangian in Eq.~(\ref{eq:yuk}) are given by
\begin{eqnarray}
	M_l &=& \left(\begin{array}{ccc}
		h_{11} \kappa' + \tilde{h}_{11} \kappa & h_{12} \kappa' + \tilde{h}_{12} \kappa & h_{13} \kappa' + \tilde{h}_{13} \kappa \\
		\frac{\kappa'^2 - \kappa^2}{\kappa'} h_{21} & h_{22} \kappa' + \tilde{h}_{22} \kappa & h_{23} \kappa' + \tilde{h}_{23} \kappa \\
		\frac{\kappa'^2 - \kappa^2}{\kappa'} h_{31} & \frac{\kappa'^2 - \kappa^2}{\kappa'} h_{32} & h_{33} \kappa' + \tilde{h}_{33} \kappa
	\end{array} \right), \\
	M_D &=& \left(\begin{array}{ccc}
		h_{11} \kappa + \tilde{h}_{11} \kappa' & h_{12} \kappa + \tilde{h}_{12} \kappa' & h_{13} \kappa + \tilde{h}_{13} \kappa' \\
		0 & h_{22} \kappa + \tilde{h}_{22} \kappa' & h_{23} \kappa + \tilde{h}_{23} \kappa' \\
		0 & 0 & h_{33} \kappa + \tilde{h}_{33} \kappa'
	\end{array} \right), \\
	M_N &=& \left(\begin{array}{ccc}
		\delta M & f_{12} v_{R1} & 0 \\
		f_{12} v_{R1} & 0 & 0 \\
		0 & 0 & 2 f_{33} v_{R2}
	\end{array} \right),
\end{eqnarray}
where $h = h^l$ and $\tilde{h} = \tilde{h}^l$. Note that a small component $\delta M$ is introduced in $M_N$ to lift the degeneracy between two RH neutrinos, and this is a crucial parameter in the calculation of the $\CP$ asymmetry in resonant leptogenesis. Introducing the short-hand notations $a \equiv h_{11} \kappa + \tilde{h}_{11} \kappa'$, $b_i \equiv  h_{i2} \kappa + \tilde{h}_{i2} \kappa'$, $c_i \equiv  h_{i3} \kappa + \tilde{h}_{i3} \kappa'$, $M_1 \equiv f_{12} v_{R,1}$ and $M_2 \equiv 2 f_{33} v_{R,2}$, and using the seesaw formula~(\ref{eq:seesaw}), we can write the light neutrino mass matrix as
\begin{eqnarray}
	M_\nu &=&
	-\left(\begin{array}{ccc}
		\frac{2 a b_1}{M_1} - \frac{\delta M b_1^2}{M_1^2} + \frac{c_1^2}{M_2} & \frac{a b_2}{M_1} - \frac{\delta M b_1 b_2}{M_1^2} + \frac{c_1 c_2}{M_2} & \frac{c_1 c_3}{M_2} \\
		\frac{a b_2}{M_1} - \frac{\delta M b_1 b_2}{M_1^2} + \frac{c_1 c_2}{M_2} &  -\frac{\delta M b_2^2}{M_1^2} + \frac{c_2^2}{M_2} & \frac{c_2 c_3}{M_2} \\
		\frac{c_1 c_3}{M_2} & \frac{c_2 c_3}{M_2} & \frac{c_3^2}{M_2}
	\end{array} \right).
	\label{eq:Mnu}
\end{eqnarray}
For an order of magnitude estimate, let us assume that each component of $M_\nu$ is of order
$\sqrt{\Delta m^2_\text{atm}} \sim 0.05$ eV. For $M_i \sim 1$ TeV and $\delta M\ll M_i$, it is clear that the natural choices are $a b_i  = (h_{11} \kappa + \tilde{h}_{11} \kappa') (h_{i2} \kappa + \tilde{h}_{i2} \kappa') \sim 10^{-8}$ GeV$^2$ and $c_i = h_{i3} \kappa + \tilde{h}_{i3} \kappa' \sim 10^{-4}$ GeV. Note that the terms $\delta M b_i b_j / M_1^2$ in Eq.~(\ref{eq:Mnu}) are negligible
as long as $|a| \gtrsim |b_i|$ and $|\delta M| \ll |M_i|$. Using these estimates, we scan over the parameter space to obtain several numerical fits to the neutrino oscillation data,
while satisfying the observed charged-lepton masses.
For illustration, we present below one such fit:
\begin{eqnarray}
	M_l &=& \left(\begin{array}{ccc}
		0.00120 & -0.0512 & -1.41 \\
		0 & 0.0926 & -0.651 \\
		0 & 0 & -0.863
	\end{array} \right) \ \text{GeV},
	\label{eq:menum}\\
	M_D &=& \left(\begin{array}{ccc}
		1.40 \ e^{0.31 \pi} & -1.55 \times 10^{-8} \ e^{-0.31 \pi} & -1.81 \times 10^{-4} \\
		0 & 2.85 \times 10^{-8} \ e^{-0.31 \pi} & -8.28 \times 10^{-5} \\
		0 & 0 & -1.11 \times 10^{-4}
	\end{array} \right) \ \text{GeV},
	\label{eq:mDnum} \\
	M_N &=& \left(\begin{array}{ccc}
		0.015 & 1752.069 & 0 \\
		1752.069 & 0 & 0 \\
		0 & 0 & -1752.044
	\end{array} \right) , \ \text{GeV}
	\label{eq:MNnum}
\end{eqnarray}
where the first two columns of $M_D$ are chosen to have opposite phases to make $M_\nu$ real
for simplicity. We will see later that these matrices correspond to the lower bound on the heavy neutrino masses that give the observed baryon asymmetry for $M_{W_R}$ = 3 TeV. We diagonalize the charged-lepton mass matrix by a bi-unitary transformation: $\widehat{M}_l = (V_l^L)^\dagger M_l V_l^R$, where
\begin{eqnarray}
	V_l^L \ = \ i \left(\begin{array}{ccc}
		-0.426 & -0.434 & -0.794 \\
		-0.236 & 0.900 & -0.366 \\
		0.874 & 0.0309 & -0.486
	\end{array} \right), \qquad
	V_l^R\  = \ i \left(\begin{array}{ccc}
		-1.00 & -0.00494 & -0.000536 \\
		-0.00494 & 1.00 & 0.00379 \\
		-0.000517 & -0.000379 & 1.00
	\end{array} \right) .
\label{Vl}
\end{eqnarray}
The light neutrino mass matrix given by Eq.~(\ref{eq:Mnu}) is diagonalized by a unitary transformation: $\widehat{M}_\nu = V_\nu^{\sf T} M_\nu V_\nu$, where
\begin{eqnarray}
V_\nu \ = \ \left(\begin{array}{ccc}
		-0.342 i & -0.0109 & 0.940 \\
		-0.745 i & 0.612 & -0.264 \\
		0.572 i & 0.790 & 0.218
	\end{array} \right) .
\end{eqnarray}
Here we have introduced a factor of $i$ into $V_\nu$ to make the light neutrino masses real and positive. In this basis, the Pontecorvo-Maki-Nakagawa-Sakata (PMNS) mixing matrix in the leptonic sector is given by
\begin{eqnarray}
	V_\text{PMNS} \ = \ (V_l^{L})^\dagger V_\nu
\ = \ \left(\begin{array}{ccc}
		0.821 & 0.551 & 0.148 \\
		-0.504 & 0.581 & 0.639 \\
		0.266 & -0.600 & 0.755
	\end{array} \right)
	\left(\begin{array}{ccc}
		1 & 0 & 0 \\
		0 & -i & 0 \\
		0 & 0 & i
	\end{array} \right), \label{pmns}
\end{eqnarray}
where the second matrix on the RHS of Eq.~(\ref{pmns}) is the Majorana phase matrix. Note that we introduced the factors of $i$ into $V_l^{L,R}$ in Eqs.~(\ref{Vl}) to write
$V_\text{PMNS}$ in its standard form (\ref{pmns}), assuming the Dirac $\CP$ phase to be zero.\footnote{In principle, one could have a fit with non-zero Dirac $\CP$ phase or with different Majorana $\CP$ phases in Eq.~(\ref{pmns}). However, this is irrelevant to our main results presented in Section~\ref{sec:3} in the sense that they simply represent viable choices of our model parameters for successful leptogenesis with low-scale $W_R$.} 
The resulting neutrino masses and mixing angles as well as the charged-lepton masses are given in Table \ref{tab:fit}. The charged-lepton masses are within 1$\%$ of their current experimental values~\cite{pdg}, and the differences can easily
be explained by electroweak radiative corrections. The mass-squared differences and mixing angles in the light neutrino sector are within 1$\sigma$ range of their global fit values~\cite{Forero:2014bxa}.
\begin{table}[t]
	\begin{center}
		\begin{tabular}{c|c} \hline \hline
			Parameter & Value \\ \hline
			$m_e$ & $0.511$ MeV  \\
			$m_\mu$ & $106$ MeV \\
			$m_\tau$ & $1.78$ GeV \\ \hline
			$m_{\nu_1}$ & $6.65 \times 10^{-3}$ eV \\
			$m_{\nu_2}$ & $1.09 \times 10^{-2}$ eV \\
			$m_{\nu_3}$ & $5.01 \times 10^{-2}$ eV \\
			\hline
			$\theta_{12}$ & $33.9^\circ$ \\
			$\theta_{23}$ & $40.3^\circ$ \\
			$\theta_{13}$ & $8.49^\circ$ \\ \hline
                        $m_{N_1}$ & $1752.044$ GeV \\
			$m_{N_2}$ & $1752.062$ GeV \\
			$m_{N_3}$ & $1752.077$ GeV \\ \hline
			\hline
		\end{tabular}
	\end{center}
\caption{The lepton masses and mixing angles calculated from the mass matrices given by Eqs.~(\ref{eq:menum})-(\ref{eq:MNnum}).}
\label{tab:fit}
\end{table}

Since $M_l = h \kappa + \tilde{h} \kappa'$ and $M_D = h \kappa' + \tilde{h} \kappa$, the Yukawa couplings $h,~\tilde{h}$ can be calculated once $\kappa$ and $\kappa'$ are appropriately chosen, with $\kappa^2+\kappa'^2=v^2$, where $v$ is the electroweak VEV. For a particular choice of
$\kappa = 112.9$ GeV and $\kappa' = 132.4$ GeV (consistent with low-energy observables and successful leptogenesis, as we will see later), we obtain
\begin{eqnarray}
	h &=& \left(\begin{array}{ccc}
		-0.0185 - 0.0273 i & 0.00142 - 3.03 \times 10^{-10} i & -0.0390 \\
		0 & 0.00256 + 5.56 \times 10^{-10} i & -0.0180 \\
		0 & 0 & -0.0239
	\end{array} \right), \label{eq:yukh} \\
	\tilde{h} &=& \left(\begin{array}{ccc}
		0.0218 + 0.0320 i & 0.00121 - 3.55 \times 10^{-10} i & 0.0333 \\
		0 & -0.00219 - 6.52 \times 10^{-10} i & 0.0154 \\
		0 & 0 & 0.0204
	\end{array} \right). \label{eq:yukht}
\end{eqnarray}

Before going into the leptogenesis analysis, we introduce an assumption that considerably simplifies the calculations. In terms of the scalar bi-doublet $\phi$ [cf.~Eq.~(\ref{eq:higgs})] and $\tilde{\phi}=\tau_2\phi^*\tau_2$, the leptonic part of the Yukawa Lagrangian (\ref{eq:yuk}) can be explicitly written as
\begin{eqnarray}
	\mathcal{L}^l_Y
%
%
	& = & h_{ij}(\phi_{1}^0 \bar{\nu}_{Li} N_{Rj} + \phi_{1}^- \bar{l}_{Li} N_{Rj} +
\phi_{2}^+ \bar{\nu}_{Li} l_{Rj} + \phi_{2}^0 \bar{l}_{Li}l_{Rj}) \nonumber \\
	&&\qquad + \: \tilde{h}_{ij}( \phi_{2}^{0*} \bar{\nu}_{Li} N_{Rj} - \phi_{2}^- \bar{l}_{Li} N_{Rj} - \phi_{1}^+ \bar{\nu}_{Li} l_{Rj} + \phi_{1}^{0*} \bar{l}_{Li}l_{Rj})+{\rm H.c.} \; ,
\end{eqnarray}
Now we assume that the doublets $\phi_1$ and $\phi_2$ are approximate mass eigenstates, and that
$\phi_1$ is heavier than the RH neutrinos which also helps suppress the flavor changing neutral currents~\cite{fcnc}. Under this assumption, not only the decay process
$N \to \phi_1 L_l$ is kinematically forbidden, but also $\phi_1$ does not contribute to the absorptive part of the one-loop self-energy correction to the decay process $N \to \phi_1 L_l$. Therefore, $\phi_1$ can be completely neglected in the $\CP$ asymmetry calculation, and $-\tilde{h}$ in Eq.~(\ref{eq:yukht}) solely determines the flavor effects relevant to the interactions
involving RH neutrinos in leptogenesis. With this assumption, $m_N$ = 1.75~TeV calculated from the example mass matrices in Eqs.~(\ref{eq:menum})-(\ref{eq:MNnum}) gives the lower bound of the RH neutrino mass for $M_{W_R}$ = 3 TeV. For higher values of $M_{W_R}$, lower $m_N$ values are possible (see Section~\ref{sec:3}, and Figure~\ref{fig:Y3} in particular).
\subsection{Predictions for Low-energy Observables} \label{sec:2b}
The class of L-R seesaw models with large Yukawa couplings gives rise to potentially large contributions to various low-energy observables~\cite{ours}, most notably lepton flavor violation (LFV) and neutrinoless double beta decay ($0\nu\beta\beta$). So we need to check whether the mass matrices given in Eqs.~(\ref{eq:menum})-(\ref{eq:MNnum}) satisfy these constraints. We start with the LFV process $\mu^- \to e^-e^+e^-$, which receives a tree-level contribution from the doubly-charged scalar fields in L-R model~\cite{pal}:
\begin{eqnarray}
	\text{BR} (\mu \to 3 e) \approx \frac{1}{2} \left(\frac{M_{W_L}}{M_{W_R}} \right)^4 \left| \frac{M_{N,12}' M_{N,11}'}{M_{\Delta_R^{++}}^2} \right|^2,
\end{eqnarray}
where $M'_{N,ij}$ are the elements of the RH neutrino mass matrix in the basis in which the charged-lepton mass matrix is diagonal. For the numerical fit presented in Section~\ref{sec:2a}, the current experimental upper limit $\text{BR} (\mu \to 3 e) < 1.0 \times 10^{-12}$~\cite{pdg} implies a lower limit of $M_{\Delta_R^{++}} \gtrsim 4$ TeV.

The scalar sector also induces one-loop contributes to the process $\mu\to e\gamma$~\cite{rnm-92}:
\begin{eqnarray}
	\text{BR} (\mu \to e \gamma)_{\Delta_R} \approx \frac{2 \alpha_W M_{W_L}^4}{3 \pi g_L^4} \left| \frac{(f'^\dagger f')_{12}}{M_{\Delta_R^{++}}^2} \right|^2 ,
\end{eqnarray}
where $f' \equiv -(V_l^R)^\mathsf{T} f V_l^R$ and $\alpha_W\equiv g_L^2/4\pi$. For the numerical fit in Section~\ref{sec:2a} and for $M_{\Delta_R^{++}}$ = 4 TeV, we obtain $\text{BR} (\mu \to e \gamma)_{\Delta_R} \simeq 1.1 \times 10^{-15}$.  There is a similar contribution from the singly-charged scalars in the loop:
\begin{eqnarray}
	\text{BR} (\mu \to e \gamma)_{\phi} \approx \frac{2 \alpha_W M_{W_L}^4}{3 \pi g_L^4} \left| \frac{(h' h'^\dagger)_{12}}{M_{\phi_1^+}^2} \right|^2 ,
\end{eqnarray}
where $h' \equiv (V_l^L)^\dagger h V_l^R$. For $M_{\phi_1^+}$ = 10 TeV, we obtain $ \text{BR} (\mu \to e \gamma)_{\phi}= 6.8 \times 10^{-18}$. There are additional contributions to $ \text{BR} (\mu \to e \gamma)$ from loops involving $W_L$~\cite{mueg-L} and $W_R$~\cite{mueg-R}, which however turn out to be sub-dominant in our model.
The total effect, including interference between the scalar loop diagrams, is found to be
$\text{BR} (\mu \to e \gamma) = 1.3 \times 10^{-15}$ for $M_{W_R}=3$ TeV, which is consistent with the current experimental limit: $\text{BR} (\mu \to e \gamma)_{\rm exp} < 5.7 \times 10^{-13}$ at 90\% CL~\cite{Adam:2013mnn}. Similar analysis can be done for $\tau \to \mu \gamma$ and  $\tau \to e \gamma$ processes; however, these LFV decays are highly suppressed due to our choice of the Yukawa coupling texture [cf.~Eqs.~(\ref{eq:yukh}) and (\ref{eq:yukht})].\footnote{Recall that this texture provided for a way to get small neutrino masses without necessarily making any of the model parameters arbitrarily small. One can of course choose different textures so that the $\tau$-induced LFV decay rates can be enhanced. Since our goal in this paper is to point out a way to relax the leptogenesis lower bound on $W_R$ mass, we only considered those textures which give an enhanced $\mu\to e\gamma$ branching ratio, which has better experimental prospects of being tested in near future.} Our results are summarized in Table~\ref{tab4}, along with the corresponding experimental limits.

Another important LFV process is the $\mu-e$ conversion, for which the dominant contribution in our model comes from the $W_L$-loops. This has been recently reevaluated in~\cite{alonso} (see also~\cite{mue-others}), and we use their expressions to compute the model predictions for three relevant nuclei, as shown in Table~\ref{tab4}.

\begin{table}[t]
  \begin{center}
    \begin{tabular}{c|c|c}\hline\hline
      Low-energy &  Model & Experimental\\
      observables & Prediction & Limit \\ \hline\hline
      BR($\mu\to e\gamma$) & $1.3\times 10^{-15}$ & $<5.7\times 10^{-13}$~\cite{Adam:2013mnn} \\
      BR($\tau \to \mu\gamma$) &  $2.4\times 10^{-17}$ & $<4.4\times 10^{-8}$~\cite{pdg} \\
      BR($\tau \to e \gamma$) & $5.7\times 10^{-17}$ & $<3.3\times 10^{-8}$~\cite{pdg} \\ \hline
      BR($\mu \to 3e$) & $9.3\times 10^{-13}$ & $<1.0\times 10^{-12}$~\cite{pdg}\\ \hline
      R$_{\mu\to e}^{\text{Ti}}$ & $2.9\times 10^{-18}$ & $< 6.1 \times 10^{-13}$~\cite{mueTi}\\
      R$_{\mu\to e}^{\text{Au}}$ & $1.0\times 10^{-17}$ & $< 7.0 \times 10^{-13}$~\cite{mueAu} \\
      R$_{\mu\to e}^{\text{Pb}}$ & $5.1\times 10^{-19}$ &  $< 4.6 \times 10^{-11}$~\cite{muePb}\\ \hline
      $T^{0\nu}_{1/2}(^{76}{\rm Ge})$ [yr] & $(0.2-1.8)\times 10^{27}$ & $> 3\times 10^{25}$ ~\cite{gerda} \\
      $T^{1/2}_{1/2}(^{136}{\rm Xe})$ [yr] & $(5.7-7.3)\times 10^{26}$ & $> 2.6\times 10^{25}$ ~\cite{kamland} \\
$T^{1/2}_{1/2}(^{130}{\rm Te})$ [yr] & $(5.7-9.7)\times 10^{25}$ & $> 2.8\times 10^{24}$ ~\cite{cuo} \\
      \hline\hline
    \end{tabular}
  \end{center}
\caption{Model predictions for  the low-energy observables with the fit shown in Section~\ref{sec:2a} and  their comparison with  the current experimental limits. }
\label{tab4}
\end{table}

For completeness, we also show the model predictions for the $0\nu\beta\beta$ half life of three relevant nuclei. Here we have taken into account the mixed LH-RH contributions~\cite{LR-mixed}, in addition to the canonical light neutrino contribution~\cite{racah} and purely RH contributions~\cite{RH}. The range of the half life prediction corresponds to the nuclear matrix element
uncertainties~\cite{NME} in various $0\nu\beta\beta$ amplitudes in the L-R model. We find that due to relatively large $W_L-W_R$ mixing in our model, $\tan{2\xi}\simeq 2\kappa\kappa'/v_R^2 \sim 7.0\times 10^{-4}$, the $\eta$-contribution is the dominant one.

It should be noted that some of our model predictions in Table~\ref{tab4} are within reach of next generation experiments, e.g. MEG-II~\cite{meg2} for BR($\mu\to e\gamma$), GERDA-II~\cite{gerda2} and Majorana~\cite{majorana} for $T^{0\nu}_{1/2}(^{76}{\rm Ge})$, EXO-200~\cite{exo} for $T^{0\nu}_{1/2}(^{136}{\rm Xe})$ and CUORE~\cite{cuore} for $T^{0\nu}_{1/2}(^{130}{\rm Te})$. Also note that we can choose different values of model parameters that give smaller LFV and $0\nu\beta\beta$ rates and
it is still possible to have the observed baryon asymmetry with $M_{W_R}$ = 3 TeV (see Section~\ref{sec:3}). Therefore, new stronger experimental constraints on low-energy observables do not necessarily require any larger $M_{W_R}$ to have successful leptogenesis.
\section{Thermodynamic evolution of lepton asymmetry in presence of Right-Handed interactions}\label{sec:3}
The time-evolution of the number density of heavy neutrinos and lepton asymmetries can be
described by a set of coupled Boltzmann equations~\cite{reviews}. Adopting the formalism of~\cite{Pilaftsis:2005rv}, we write down the flavor-diagonal Boltzmann equations in terms of the normalized number densities of heavy neutrinos $\eta_\alpha^N=n^N_\alpha/n_\gamma$ and the normalized lepton asymmetries $\eta^{\Delta L}_l = (n^L_l-\bar{n}^L_l)/n_\gamma$:
\begin{align}
\frac{H_N n_\gamma}{z}\frac{d\eta^N_\alpha}{dz} \ & = \ -\left(\frac{\eta^N_\alpha}{\eta^N_{\rm eq}}-1\right)\sum_k \left(\gamma^D_{k\alpha} + \gamma^{S_L}_{k\alpha}+\gamma^{S_R}_{k\alpha}\right) \;, \label{be1} \\
\frac{H_N n_\gamma}{z}\frac{d\eta^{\Delta L}_l}{dz} \ & = \ \sum_\alpha \varepsilon_{l\alpha}\left(\frac{\eta^N_\alpha}{\eta^N_{\rm eq}}-1\right)\sum_k \tilde{\gamma}^D_{k\alpha}
\nonumber \\
\ & \
\qquad
-\frac{2}{3}\eta^{\Delta L}_l\left[\sum_\alpha \left(B_{l\alpha}\sum_k {\gamma}^D_{k\alpha}+\tilde{\gamma}^{S_L}_{l\alpha}+\tilde{\gamma}^{S_R}_{l\alpha}
\right)
+ \sum_k \left(\gamma^{(\Delta L=2)}_{lk} + \gamma^{(\Delta L=0)}_{lk} \right)
\right] \;, \label{be2}
\end{align}
where $k,l = e,\mu,\tau$ and $\alpha=1,2,3$ are the lepton and heavy neutrino indices respectively, $z=m_{N_1}/T$ is a dimensionless variable, $H_N\equiv H(z=1)\simeq 17 m_{N_1}^2/M_{\rm Pl}$ is the Hubble parameter at $z=1$, assuming only SM degrees of freedom in the thermal bath. The number densities of heavy neutrinos and lepton asymmetries are normalized to the photon number density
\begin{eqnarray}
n_\gamma \ = \ \frac{2m_{N_1}^3\zeta(3)}{\pi^2 z^3} \; ,
\end{eqnarray}
where $\zeta(x)=\displaystyle{\sum_{n=1}^\infty} n^{-x}$ is the Riemann zeta function, with $\zeta(3)\approx 1.20206$. The normalized equilibrium number density of the heavy neutrinos is given by
\begin{eqnarray}
\eta^N_{\rm eq} \ \equiv \ \frac{n^N_{\rm eq}}{n_\gamma} \ = \ \frac{1}{2\zeta(3)}z^2 K_2(z) \; ,
\label{etaeq}
\end{eqnarray}
where $K_n(x)$ is the $n$th-order modified Bessel function of the second kind.

In Eq.~(\ref{be2}), $\varepsilon_{l\alpha}$ and $B_{l\alpha}$ are respectively the individual lepton-flavor $\CP$ asymmetries and branching ratios for the heavy Majorana neutrino decays generating the lepton number asymmetry:
\begin{align}
\varepsilon_{l\alpha} \ & = \ \frac{1}{\Gamma_{N_\alpha}}\left[\Gamma(N_\alpha \to L_l \phi) - \Gamma(N_\alpha\to L_l^c\phi^c)\right], \label{eps} \\
B_{l\alpha} \ & = \ \frac{1}{\Gamma_{N_\alpha}}\left[\Gamma(N_\alpha \to L_l \phi) + \Gamma(N_\alpha\to L_l^c\phi^c) \right]. \label{BR}
\end{align}
The partial decay widths in Eqs.~(\ref{eps}) and (\ref{BR}) are expressed as
\begin{equation}
  \Gamma(N_{\alpha}\to L_l\phi) \
  = \ m_{N_\alpha}A^l_{\alpha\alpha}(\widehat{\maf{h}})\;,
  \qquad \qquad
  \Gamma(N_{\alpha} \to L_l^c\phi^\dag) \
  = \ m_{N_\alpha}A^l_{\alpha\alpha}(\widehat{\maf{h}}^c)\;,
  \label{gamma}
\end{equation}
where $A^l_{\alpha\beta}$ are the absorptive transition amplitudes:
\begin{equation}
  A^l_{\alpha\beta}(\widehat{\maf{h}}) \
    = \ \frac{1}{16\pi}{\widehat{\maf{h}}}_{l \alpha}{\widehat{\maf{h}}}^*_{l \beta} \;,
\label{Aab}
\end{equation}
and $\widehat{\maf{h}}$ are the one-loop resummed effective Yukawa couplings~\cite{Pilaftsis:2003gt}, which take into account the unstable particle-mixing effects in the heavy neutrino self-energy diagrams in the resonant leptogenesis scenario. In the heavy neutrino mass eigenbasis, we have~\cite{Pilaftsis:2003gt, Pilaftsis:2008qt}
\begin{align}
  &{\widehat{\maf{h}}}_{l \alpha} \ = \ \widehat{h}_{l \alpha} \:
  - \: i\sum_{\beta,\gamma}|\epsilon_{\alpha\beta\gamma}|\widehat{h}_{l \beta}
  \label{resum1}\\
  & \ \times
\frac{m_\alpha(m_\alpha A_{\alpha\beta} + m_\beta A_{\beta\alpha})
    \: - \: iR_{\alpha\gamma}[m_\alpha A_{\gamma\beta}(m_\alpha A_{\alpha\gamma}
    + m_\gamma A_{\gamma\alpha})\:
    + m_\beta A_{\beta\gamma}(m_\alpha A_{\gamma\alpha} \:
    + m_\gamma A_{\alpha\gamma})]}
  {m^2_\alpha \: - \: m^2_\beta \: + \: 2i m^2_\alpha A_{\beta\beta} \:
    + \: 2i {\rm Im}(R_{\alpha \gamma})
    [m^2_\alpha|A_{\beta\gamma}|^2 \:
    + \: m_\beta m_\gamma {\rm Re}(A^2_{\beta\gamma})]}\;,
  \nonumber
\end{align}
where   $\epsilon_{\alpha\beta\gamma}$   is   the  usual   Levi-Civita
antisymmetric tensor, and
\begin{equation}
  R_{\alpha\beta} \ = \ \frac{m_\alpha^2}
  {m_\alpha^2 \: - \: m_\beta^2 \: + \: 2im_\alpha^2 A_{\beta\beta}}\; .
  \label{Rab}
\end{equation}
All   the  transition   amplitudes
$A_{\alpha\beta}   \:   \equiv  \:
A_{\alpha\beta}(\widehat{h})$ in Eq.~(\ref{resum1})  are evaluated on-shell with $p^2
  =   m^2_{N_\alpha}    \equiv    m^2_\alpha$ using the tree-level Yukawa
couplings $\widehat{h}_{l\alpha}$. The  $ \CP$-conjugate      effective      Yukawa      couplings ${\maf{h}}^c_{l\alpha}$ can be obtained from Eq.~(\ref{resum1}) by
replacing  $h_{l\alpha}$  with their complex  conjugates  $h^*_{l\alpha}$.
Using Eqs.~(\ref{gamma}) and
summing over all charged-lepton flavors, we obtain the total 2-body decay
width
\begin{equation}
\Gamma^{N_\alpha}_{L\phi} \ = \ \sum_l \left[\Gamma(N_{\alpha}\to L_l\phi)+\Gamma(N_{\alpha}\to L_l^c\phi^c)\right]
\ = \
\frac{m_{N_\alpha}}{16 \pi}\,\left[ (\mathbf{\widehat{h}}^{\dagger} \,  \mathbf{\widehat{h}})_{\alpha \alpha}  \; + \; (\mathbf{\widehat{h}}^{c\dagger} \, \mathbf{\widehat{h}}^c)_{\alpha \alpha} \right].
\label{resum_width0}
\end{equation}

In addition to the usual 2-body decay mode discussed above, the L-R model gives rise to a 3-body decay mode of $N_\alpha$, mediated by an off-shell $W_R$. This is a $C\!P$-conserving channel, being governed only by gauge interactions, and does not lead to any additional lepton asymmetry, but only contributes to the depletion of $N_\alpha$ and dilution of the lepton asymmetry.  Assuming $m_{N_\alpha}\leq M_{W_R}$,\footnote{Note that for $m_{N_\alpha}> M_{W_R}$, the decay mode $N\to l_RW_R$ will become dominant over the $N\to L\phi$ mode, which makes leptogenesis unfeasible~\cite{Carlier:1999ac}. Therefore, we will only consider the case with $m_{N_\alpha}\leq M_{W_R}$.} the 3-body decay width is given by
\begin{eqnarray}
\Gamma(N_\alpha\to l_R q_R\bar{q}'_R) \ = \ \Gamma(N_\alpha\to \bar{l}_R \bar{q}_R q'_R) \ = \ \frac{3g_R^4}{2^9\pi^3 m_{N_\alpha}^3} \int_0^{m_{N_\alpha}^2} ds \frac{m_{N_\alpha}^6-3m_{N_\alpha}^2 s^2+2 s^3}{(s-M_{W_R}^2)^2+M_{W_R}^2\Gamma_{W_R}^2} \; ,
\label{3body}
\end{eqnarray}
where $\Gamma_{W_R}\simeq (g_R^2/4\pi)M_{W_R}$ is the total decay width of $W_R$, assuming that all three heavy neutrinos are lighter than $W_R$.
Thus, the {\it total} heavy neutrino decay width appearing in the denominator of Eqs.~(\ref{eps}) and (\ref{BR}) is given by
\begin{eqnarray}
\Gamma_{N_\alpha} \ = \ \sum_l\left[\Gamma(N_\alpha \to L_l \phi) + \Gamma(N_\alpha\to L_l^c\phi^\dag) \right] + 2\: \Gamma(N_\alpha\to l_R q_R\bar{q}'_R) \; .
\label{resum_width}
\end{eqnarray}

Using the above definitions, the flavored $\CP$-asymmetries $\varepsilon_{l\alpha}$ for the example fit in Section~\ref{sec:2a} are given by
\begin{eqnarray}
\varepsilon = \left(\begin{array}{ccc}
-0.103882 & -0.528076 & -0.0699693\\
-0.0171983 & -0.0069479 & -0.0162488\\
0 & 0 & 0
\end{array}\right) . \label{cpasy}
\end{eqnarray}
Thus, in our model,  the $\CP$-asymmetries in the electron and muon sector could be resonantly enhanced and become close to maximal, while there is no $\CP$-asymmetry generated in the tau-sector. Thus, the final lepton asymmetry will only be generated in the electron and muon sectors, as we discuss below.
\subsection{Decay and Scattering Terms}
The various decay and scattering rates in Eqs.~(\ref{be1}) and (\ref{be2}) are given below in terms of the physical decay and scattering parameters involving the heavy neutrinos:\footnote{Here we have ignored the sub-dominant chemical potential contributions from the lepton, quark and Higgs fields~\cite{Pilaftsis:2005rv}.}
\begin{align}
\gamma^D_{l\alpha} \ & = \ \gamma^{N_\alpha}_{L_l\phi_l} + \gamma^{N_\alpha}_{l_Rq\bar{q}'},
\label{decay} \\
\tilde{\gamma}^D_{l\alpha} \ & = \ \gamma^{N_\alpha}_{L_l\phi_l},\\
\gamma^{S_L}_{l\alpha} \ & = \ \gamma^{N_\alpha L_l}_{Qu^c}+\gamma^{N_\alpha u^c}_{L_l Q^c} + \gamma^{N_\alpha Q}_{L_l u} + \gamma^{N_\alpha L_l}_{\phi^\dag V_\mu} + \gamma^{N_\alpha V_\mu}_{L_l \phi} + \gamma^{N_\alpha \phi^\dag}_{L_l V_\mu}, \\
\tilde{\gamma}^{S_L}_{l\alpha} \ & = \ \frac{\eta^N_\alpha}{\eta^N_{\rm eq}}\gamma^{N_\alpha L_l}_{Qu^c}+\gamma^{N_\alpha u^c}_{L_l Q^c} + \gamma^{N_\alpha Q}_{L_l u} + \frac{\eta^N_\alpha}{\eta^N_{\rm eq}}\gamma^{N_\alpha L_l}_{\phi^\dag V_\mu} + \gamma^{N_\alpha V_\mu}_{L_l \phi} + \gamma^{N_\alpha \phi^\dag}_{L_l V_\mu}, \\
\gamma^{S_R}_{l\alpha} \ & = \ 
\gamma^{N_\alpha l_R}_{\bar{u}_R d_R} + \gamma^{N_\alpha \bar{u}_R}_{l_R \bar{d}_R} + \gamma^{N_\alpha d_R}_{l_R u_R},\\
 \tilde{\gamma}^{S_R}_{l\alpha} \ & = \ 
\frac{\eta^N_\alpha}{\eta^N_{\rm eq}} \gamma^{N_\alpha l_R}_{\bar{u}_R d_R} + \gamma^{N_\alpha \bar{u}_R}_{l_R \bar{d}_R} + \gamma^{N_\alpha d_R}_{l_R u_R},\\
\gamma^{(\Delta L=2)}_{lk} \ & = \ \gamma'^{L_l\phi_l}_{L_k^c\phi_k^\dag}+\gamma^{L_l L_k}_{\phi_l^\dag \phi_k^\dag}, \\
\gamma^{(\Delta L=0)}_{lk} \ & = \ \gamma'^{L_l\phi_l}_{L_k\phi_k}+\gamma^{L_l\phi_l^\dag}_{L_k\phi_k^\dag}+\gamma^{L_l L_k^c}_{\phi_l \phi_k^\dag} \; .
\end{align}
The scattering terms involving two heavy neutrinos in the initial state, e.g. induced by a $t$-channel $W_R$ or $e_R$, and by an $s$-channel $Z_R$, are not included here since their rates are doubly Boltzmann-suppressed and numerically much smaller than the scattering rates given above~\cite{hambye, Blanchet:2009bu, Blanchet:2010kw}.\footnote{In a version of the L-R model where the
$\phi_l$'s are leptophilic~\cite{ours}, the $\Delta L=1$ scatterings involving SM quarks, e.g. $N_\alpha L_l\leftrightarrow Qu^c $, are also absent.}

The decay rates are explicitly given by
\begin{align}
\gamma^{N_\alpha}_{L_l\phi} \ & = \ \frac{m_{N_\alpha}^3}{\pi^2 z}K_1(z)\left[\Gamma(N_\alpha\to L_l \phi)+\Gamma(N_\alpha\to L_l^c\phi^\dag)\right], \\
\gamma^{N_\alpha}_{l_Rq\bar{q}'} \ & = \ \frac{m_{N_\alpha}^3}{\pi^2 z} K_1(z)\left[\Gamma(N_\alpha\to l_R q_R\bar{q}'_R)+\Gamma(N_\alpha\to \bar{l}_R\bar{q}_R q'_R)\right] \; .
\end{align}
The various collision terms for the $2\leftrightarrow 2$ scattering processes $XY\leftrightarrow AB$ are defined as
\begin{eqnarray}
\gamma^{XY}_{AB} \ = \ \frac{m_{N_1}^4}{64\pi^4 z}\int _{x_{\rm thr}}^\infty dx\sqrt{x} K_1(z\sqrt{x})\hat{\sigma}^{XY}_{AB}(x), \label{scat}
\end{eqnarray}
where $x=s/m_{N_1}^2$ with the kinematic threshold value $x_{\rm thr} = {\rm max}[(m_X+m_Y)^2,(m_A+m_B)^2]/m_{N_1}^2$, and $\hat{\sigma}^{XY}_{AB}(x)$ are the relevant reduced cross sections, whose explicit expressions are given in Appendix~\ref{app:A}.

For $\Delta L=2$ processes $L_l\phi \leftrightarrow L^c_k\phi^c$ and $L_lL_k\leftrightarrow \phi^c \phi^c$, and $\Delta L=0$ processes $L_l\phi \leftrightarrow L_k \phi$, $L_l\phi^c \leftrightarrow L_k \phi^c$, $L_l L_k^c \leftrightarrow \phi\phi^c$, only the resonant parts of $L_l\phi \leftrightarrow L^c_k\phi^c$ and $L_l\phi \leftrightarrow L_k \phi$ give the dominant contributions in the resonant leptogenesis scenario.  Using the narrow width approximation for the resummed heavy-neutrino propagators, these scattering rates can be written as~\cite{Deppisch:2010fr}
\begin{align}
 \gamma^{L_k\Phi}_{L_l\Phi} & \ = \ \sum_{\alpha,\beta}
  \frac{\left(\gamma^{N_\alpha}_{L\Phi}+
      \gamma^{N_\beta}_{L\Phi}\right)}
  {\left(1-2i \: \frac{m_{N_\alpha}\: - \: m_{N_\beta}}
      {\Gamma_{N_\alpha} \: + \: \Gamma_{N_\beta}}\right)}
  \frac{2\left(\widehat{\maf{h}}^*_{l\alpha} \widehat{\maf{h}}^{c^*}_{k\alpha}
      \widehat{\maf{h}}_{l\beta}\widehat{\maf{h}}^c_{k\beta}
      \: + \: \widehat{\maf{h}}^{c^*}_{l\alpha}\widehat{\maf{h}}^*_{k\alpha}
      \widehat{\maf{h}}^c_{l\beta}\widehat{\maf{h}}_{k\beta}\right)}
  {\left[(\widehat{\maf{h}}^\dag\widehat{\maf{h}})_{\alpha\alpha}
      + (\widehat{\maf{h}}^{c^\dag}\widehat{\maf{h}}^c)_{\alpha\alpha}
      + (\widehat{\maf{h}}^\dag\widehat{\maf{h}})_{\beta\beta}
      + (\widehat{\maf{h}}^{c^\dag}\widehat{\maf{h}}^c)_{\beta\beta} \right]^2}\;,
  \label{scat1}\\
  \gamma^{L_k\Phi}_{L^c_l\Phi^c} & \ = \ \sum_{\alpha,\beta}
  \frac{\left(\gamma^{N_\alpha}_{L\Phi}+
      \gamma^{N_\beta}_{L\Phi}\right)}
  {\left(1-2i \: \frac{m_{N_\alpha}\: - \: m_{N_\beta}}
      {\Gamma_{N_\alpha}\: + \: \Gamma_{N_\beta}}\right)}
  \frac{2\left(\widehat{\maf{h}}^*_{l\alpha} \widehat{\maf{h}}^*_{k\alpha}
      \widehat{\maf{h}}_{l\beta}\widehat{\maf{h}}_{k\beta}
      \: + \: \widehat{\maf{h}}^{c^*}_{l\alpha}\widehat{\maf{h}}^{c^*}_{k\alpha}
      \widehat{\maf{h}}^c_{l\beta}\widehat{\maf{h}}^c_{k\beta}\right)}
  {\left[(\widehat{\maf{h}}^\dag\widehat{\maf{h}})_{\alpha\alpha}
      + (\widehat{\maf{h}}^{c^\dag}\widehat{\maf{h}}^c)_{\alpha\alpha}
      + (\widehat{\maf{h}}^\dag\widehat{\maf{h}})_{\beta\beta}
      + (\widehat{\maf{h}}^{c^\dag}\widehat{\maf{h}}^c)_{\beta\beta}
    \right]^2}\; ,\label{scat2}
\end{align}
where $\gamma^{N_\alpha}_{L\phi}\equiv \sum_l \gamma^{N_\alpha}_{L_l\phi}$, and $\Gamma_{N_\alpha}$ is the total decay width of $N_\alpha$ given by (\ref{resum_width}). The  RIS-subtracted collision rates $\gamma'^{L_k\phi}_{L_l\phi}$ and $\gamma'^{L_k\phi}_{L^c_l\phi^c}$ can be obtained from Eqs.~(\ref{scat1}) and (\ref{scat2}) respectively by taking $\alpha\neq \beta$.
\subsection{Analytic Solution for Lepton Asymmetry}
\begin{figure}[t]
\includegraphics[width=14cm]{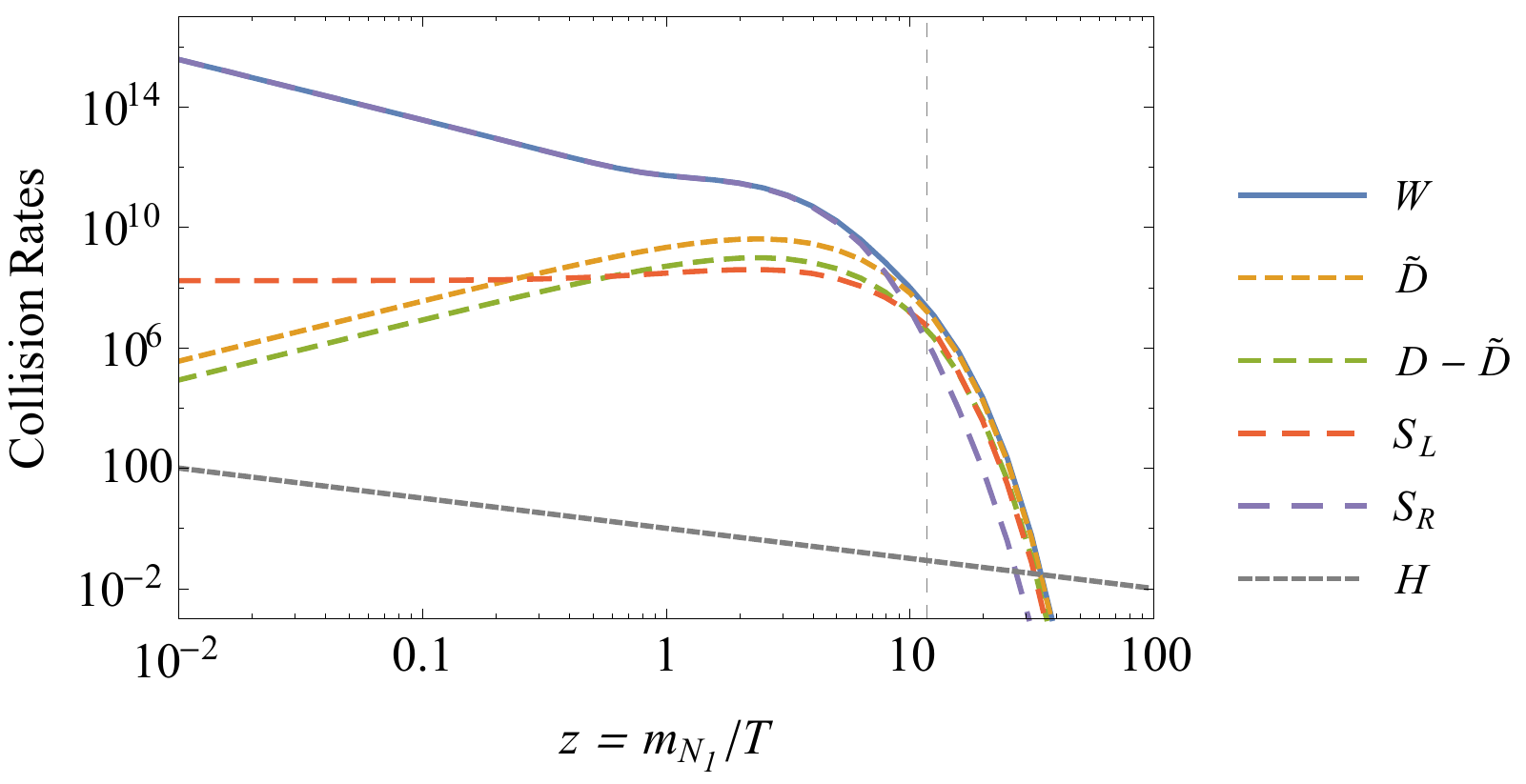}
\caption{The dimensionless collision rate parameters defined in Eqs.~(\ref{D})-(\ref{W}), compared with the rescaled Hubble rate $H=zH(z)/H_N$. We have shown the results for $\alpha=1$ and $l=1$, and for the fit given in Section~\ref{sec:2a}. The vertical dashed line shows the critical temperature $z_c=m_{N_1}/T_c$, beyond which the sphaleron transitions freeze out.} \label{fig:rate}
\end{figure}

Using Eq.~(\ref{be1}),  Eq.~(\ref{be2}) can be rewritten as
\begin{eqnarray}
\frac{d\eta^{\Delta L}_{l}}{dz} \ & = & \ - \sum_\alpha \varepsilon_{l\alpha}\frac{d\eta^N_{\alpha}}{dz} \frac{\tilde{D}_{\alpha}}{D_\alpha+S^L_\alpha+S^R_\alpha}- \frac{2}{3}\eta^{\Delta L}_{l} W_l(z) \; , \label{be3}
\end{eqnarray}
\begin{eqnarray}
{\rm where}\quad \tilde{D}_\alpha & = & \frac{z}{H_N n^\gamma}\sum_k \tilde{\gamma}^D_{k\alpha} \; , \label{D}\\
D_\alpha & = & \frac{z}{H_N n^\gamma}\sum_k \gamma^D_{k\alpha} \; ,\\
S^{L,R}_\alpha & = & \frac{z}{H_N n^\gamma} \sum_k \gamma^{S_{L,R}}_{k\alpha} \;, \\
W_l & = & \frac{z}{H_Nn^\gamma} \left[\sum_\alpha \left(B_{l\alpha}\sum_k {\gamma}^D_{k\alpha}+\tilde{\gamma}^{S_L}_{l\alpha}+\tilde{\gamma}^{S_R}_{l\alpha}\right) + \sum_k \left(\gamma^{(\Delta L=2)}_{lk} + \gamma^{(\Delta L=0)}_{lk} \right) \right]  \; . \label{W}
\label{be42}
\end{eqnarray}
For the example fit discussed in Section~\ref{sec:2a}, the dimensionless collision terms given above are shown numerically in Figure~\ref{fig:rate}. Here $H=zH(z)/H_N$ is a measure of the
Hubble expansion rate, $\tilde{D}~(D)$ denotes the 2-body (total) decay rate of $N$, $D-\tilde{D}$ denotes the 3-body decay rate of $N$, $S_L$ and $S_R$ are the scattering rates induced by LH and RH currents respectively and $W$ is the total washout rate. The vertical dashed line shows the critical temperature $z_c=m_{N_1}/T_c$, where $T_c$ is the critical temperature for the electroweak phase transition, given at one loop by~\cite{Cline:1993bd}
\begin{eqnarray}
T_c^2 \ = \ \frac{1}{4D}\left[M_H^2-\frac{3}{8\pi^2 v^2}(2M_W^4+M_Z^4-4M_t^4)-\frac{1}{8\pi^2v^4D}(2M_W^3+M_Z^3)^2\right] \; , \label{tc}
\end{eqnarray}
with $D=\frac{1}{8v^2}(2M_W^2+M_Z^2+2m_t^2+M_H^2)$. For $T<T_c$ (or equivalently, for $z>z_c$),
the conversion of lepton asymmetry to the baryon sector freezes out as the sphaleron processes become ineffective. Using the latest experimental values of the SM mass parameters, we obtain $T_c =  149.4^{+0.7}_{-0.8}~{\rm GeV}$~\cite{DMPT}. It is clear that for $z<z_c$, we are in the strong washout regime ($\Gamma \gg H$). More importantly, due to the relatively large Yukawa couplings in our model, the decay rate is sizable compared to the scattering and washout rates around $z=z_c$, thus yielding a successful leptogenesis, even with a low-scale $M_{W_R}$, as shown below.

Assuming negligible initial lepton asymmetry, the solution of Eq.~(\ref{be3}) can be written in terms of the $\CP$ asymmetry and the efficiency factor, as follows:
\begin{eqnarray}
\eta_l^{\Delta L}(z) \ = \ - \sum_\alpha \varepsilon_{l\alpha}\kappa_{l\alpha}(z) \; ,
\label{etaL0}
\end{eqnarray}
 where the efficiency factor is given by
\begin{eqnarray}
\kappa_{l\alpha}(z) &=& \int_{z_i}^z dz' \frac{d\eta^N_\alpha}{dz'}\frac{\tilde{D}_\alpha}{D_\alpha+S_\alpha}{\rm exp}\left[-\frac{2}{3}\int_{z'}^{z} dz'' W_l(z'')\right] \; , \label{kappa}
\end{eqnarray}
with $S_\alpha\equiv S^L_\alpha+S^R_\alpha$ denoting the total scattering rate.
The double integral in Eq.~(\ref{kappa}) can be solved numerically to get an exact value of the lepton number asymmetry. In the strong washout regime of resonant leptogenesis, the density of heavy neutrinos closely follows its equilibrium abundance, i.e.  $(\eta^N_\alpha/\eta^N_{\rm eq}-1)\ll 1$. In this case, we can obtain an analytic solution to the integral equation~(\ref{etaL0}), following Ref.~\cite{Buchmuller:2004nz}. Using Eq.~(\ref{etaeq}), we obtain
\begin{eqnarray}
\frac{d\eta^N_\alpha}{dz} \ \simeq \ \frac{d\eta^N_{\rm eq}}{dz} \ = \ -\frac{1}{2\zeta(3)}z^2 K_1(z)\; .
\end{eqnarray}
where we have used the property of the Bessel function: $K_{n+1}(z)-K_{n-1}(z)=(2n/z)K_n(z)$. Also the inverse decay term on the RHS of Eq.~(\ref{W}) can be written as
\begin{eqnarray}
W_l \ \supset \ \frac{z}{H_Nn^\gamma}\sum_\alpha B_{l\alpha} \gamma^{N_\alpha}_{L\phi} \ = \ \frac{1}{2\zeta(3)}z^3K_1(z)\sum_\alpha B_{l\alpha} K_\alpha \ \equiv \ \frac{1}{2\zeta(3)}z^3K_1(z) K_l \; ,
\label{W1}
\end{eqnarray}
where $K_\alpha = \Gamma^{N_\alpha}_{L\phi}/H_N$. The scattering terms in Eq.~(\ref{W}) can be included in Eq.~(\ref{W1}) by scaling $K_l\to K^{\rm eff}_l = \kappa'_l K_l$~\cite{Deppisch:2010fr}, where
\begin{eqnarray}
\kappa'_l \ = \ 1+\frac{\sum_\alpha \tilde{\gamma}^S_{l\alpha}+\sum_k\left(\gamma^{(\Delta L=2)}_{lk} + \gamma^{(\Delta L=0)}_{lk} \right)}{\sum_\alpha B_{l\alpha}\gamma^{N_\alpha}_{L\phi}} \; .
\end{eqnarray}
With these substitutions, Eq.~(\ref{be3}) can be rewritten as
\begin{eqnarray}
\frac{d\eta^{\Delta L}_{l}}{dz} \ = \ \frac{1}{2\zeta(3)}z^2K_1(z)\left[\sum_\alpha \varepsilon_{l\alpha}\frac{\tilde{D}_{\alpha}}{D_\alpha+S_\alpha}- \frac{2}{3}zK^{\rm eff}_l\eta^{\Delta L}_{l}\right] \; .
\label{be4}
\end{eqnarray}
Eq.~(\ref{be4}) is an ordinary differential equation of the form $\frac{dy}{dx}+P(x)y=Q(x)$ which can be analytically solved using the integrating factor method. In the regime $2(K^{\rm eff}_l)^{-1/3} \lesssim z \lesssim 1.25\ln(25K_l^{\rm eff})\equiv z_f$, the solution to Eq.~(\ref{be4}) can be approximated by
\begin{eqnarray}
\eta^{\Delta L}_l (z) \ \simeq \ \frac{3}{2zK_l^{\rm eff}}\sum_\alpha \varepsilon_{l\alpha}\frac{\tilde{D}_{\alpha}}{D_\alpha+S_\alpha} \; .
\label{etaL}
\end{eqnarray}
Note that the final lepton asymmetry relevant for the observed baryon asymmetry should be evaluated  at $z=z_c$ as a function of the model parameters, most importantly the Yukawa couplings and $M_{W_R}$. This value should be compared with $\eta^{\Delta L}_{\rm obs} = -(2.47\pm 0.03)\times 10^{-8}$~\cite{DMPT} in order to be compatible with the $68\%$ CL Planck value for the observed baryon asymmetry in our Universe~\cite{planck}, after taking into account the sphaleron transition and entropy dilution effects. Thus for $z_f\gtrsim z_c$, Eq.~(\ref{etaL}) gives a good approximation for $\eta^{\Delta L}_l(z_c)$, irrespective of the initial conditions.

For illustration, we give below the numerical values of the final lepton asymmetry in each flavor using Eq.~(\ref{etaL}), corresponding to the fit presented in Section~\ref{sec:2a} with the $\CP$ asymmetry given by Eq.~(\ref{cpasy}):
\begin{eqnarray}
\eta^{\Delta L}_e = -2.28852\times 10^{-8}, \qquad
\eta^{\Delta L}_\mu = -1.739\times 10^{-9}, \qquad \eta^{\Delta L}_{\tau} = 0 \; . \label{flavasy}
\end{eqnarray}
Thus, the final lepton asymmetry is mostly generated in the electron sector for this particular fit, due to the structure of the Yukawa couplings in Eq.~(\ref{eq:yukht}), and of the $\CP$-asymmetry in Eq.~(\ref{cpasy}).
\subsection{Lower Bound on the Mass of $W_R$}
Before analyzing the dependence of the total lepton asymmetry $\eta^{\Delta L}(z_c)=\sum_l \eta_l^{\Delta L}(z_c)$ [cf.~Eq.~(\ref{etaL})] on the $W_R$ mass, it is useful to study its parametric dependence on the overall scale of the largest Yukawa coupling in the model, which can be schematically written as
\begin{eqnarray}
\eta^{\Delta L}(z_c) \ \simeq \ \frac{aY^2}{(a'Y^2+b')(aY^2+b)},
\label{etaL2}
\end{eqnarray}
since $\tilde{D}\sim aY^2$, while $D+S \sim aY^2+b$, and $K_l^{\rm eff}\sim a'Y^2+b'$, where $a,b,a',b'$ are $W_R$-mass dependent parameters, but independent of the Yukawa couplings. Here we have assumed that for any given Yukawa coupling matrix, the correct $\CP$-asymmetry $\varepsilon$ in Eq.~(\ref{etaL}), as required to match  $\eta^{\Delta L}_{\rm obs}$, can be obtained by changing the degeneracy parameter $\delta M$ in Eq.~(\ref{eq:texture}) and the relevant $\CP$ phases appropriately. Therefore, we have treated $\varepsilon$ as a constant parameter with respect to the scale of Yukawa couplings to write down the relation (\ref{etaL2}). From this relation, we observe that for very small Yukawa couplings, the 2-body decay rate $\tilde{D}$ is small, while the scattering effects induced by the RH gauge currents for a relatively low $W_R$-scale, give a large contribution to the dilution factor $D+S$ and the washout factor $K_l^{\rm eff}$, both of which are almost independent of $Y$ in the small Yukawa regime. Therefore, the final lepton asymmetry will be suppressed for small Yukawa couplings, as expected for vanilla seesaw, which gives the lower bound of 18 TeV on the $W_R$ mass~\cite{hambye}. On the other hand, for large Yukawa couplings, as in our model, the 2-body decay rate is large compared to the scattering and washout rates induced by RH currents, and hence, we would naively expect the lower bound on $M_{W_R}$ to be weaker. As we show below, this is indeed the case, but at the same time, we should keep in mind that the washout due to inverse decay $L\phi\to N$ as well as the $\Delta L=2$ scatterings are also large, being induced by the same large Yukawa couplings.\footnote{The 3-body inverse decay $l_Rq_R\bar{q}'_R\to N$ is sub-dominant compared to $L\phi\to N$, and can be ignored in this discussion.}  Thus, the final lepton asymmetry becomes proportional to $Y^{-2}$ in the large Yukawa limit, and therefore, the lower limit on $W_R$ cannot be  arbitrarily weakened by just increasing the scale of the Yukawa coupling matrix.
\begin{figure}[t]
\centering
\includegraphics[width=15cm]{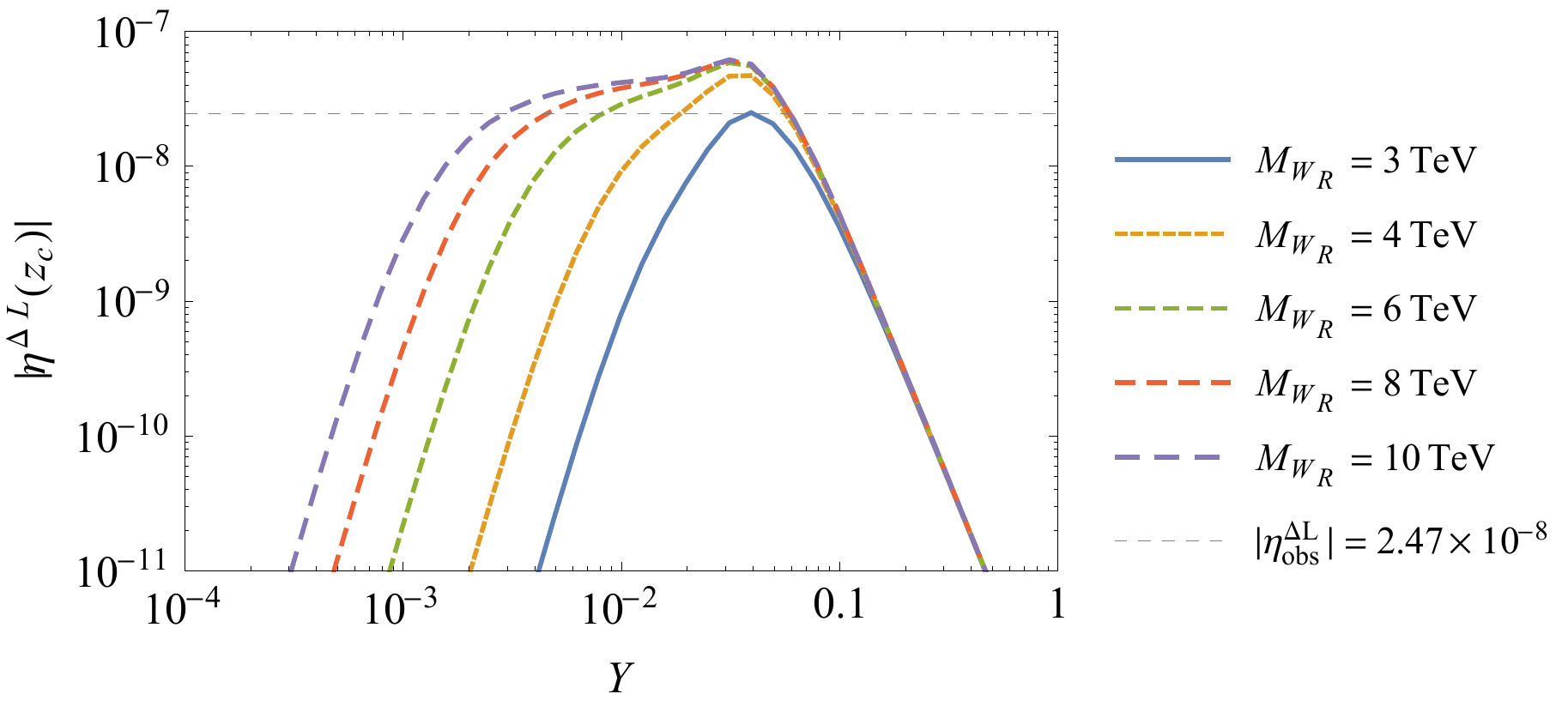}
\caption{The final lepton asymmetry as a function of the scale of Yukawa coupling matrix for various $W_R$ masses.  The dashed horizontal line shows the value required to satisfy the observed baryon asymmetry. }\label{fig:Y}
\end{figure}
\begin{figure}[h]
\centering
\includegraphics[width=10cm]{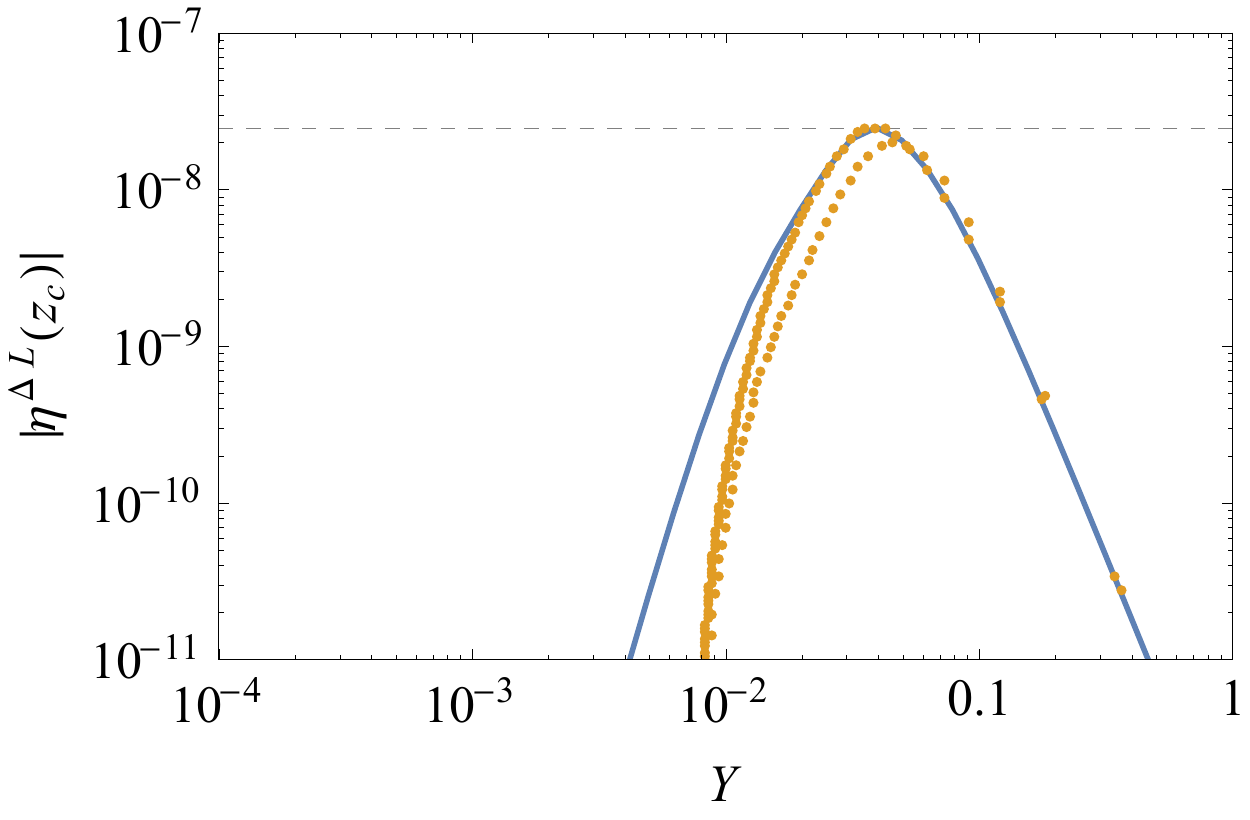}
\caption{Realistic fits shown by the dots, compared with the value obtained by rescaling the Yukawa couplings (solid line) for the $M_{W_R}=3$ TeV.}
\label{fig:Y2}
\end{figure}

From Eq.~(\ref{etaL2}) and the subsequent discussion above, it is clear that there exists some intermediate value of $Y$ for which the final lepton asymmetry is the maximum, and it decreases rapidly when either $Y\ll 1$ or $Y\to 1$. This is numerically verified in Figure~\ref{fig:Y} for various $W_R$ masses. Here $Y$ denotes the largest entry in the rescaled Yukawa coupling matrix obtained by multiplying an overall factor to Eq.~(\ref{eq:yukht}), whose flavor structure is fixed by the neutrino fit discussed in Section 2. We find from Figure~\ref{fig:Y} that there exists a range of $Y$ between $10^{-3}$ - $10^{-1}$ where the maximum lepton asymmetry is obtained in our low-scale L-R seesaw model.  This demonstrates that $M_{W_R}$ as low as 3 TeV is still compatible with successful leptogenesis, as long as the Yukawa couplings are in the favorable range.

One might question whether it is always possible to have a realistic fit with variations in the overall scale of the Yukawa coupling matrix, as done in Figure~\ref{fig:Y}. To answer this question, we present a set of realistic fits (shown by the dots) in Figure~\ref{fig:Y2} for a fixed $M_{W_R}=3$ TeV. For comparison, we also show the result obtained by  varying the VEVs $\kappa,~\kappa'$ and the Yukawa couplings (solid line, same as in Figure~\ref{fig:Y} for $M_{W_R}=3$ TeV) for fixed mass matrices given by Eqs.~(\ref{eq:menum})-(\ref{eq:MNnum}), while requiring them to satisfy all the low-energy constraints, as discussed in Section~\ref{sec:2b}. We find that the two results are quite close for $Y\gtrsim 10^{-2}$. There appear two branches of solutions around the maximum value because two lepton asymmetry values exist around $\kappa = \kappa'$ which is the singular point where we have no solution of Yukawa couplings for given mass matrices Eqs.~(\ref{eq:menum})-(\ref{eq:MNnum}). For smaller values of $Y$, the lepton asymmetry decreases much faster than that expected from Eq.~(\ref{etaL2}), presumably because the $\CP$-asymmetry depends on $Y$ in this regime.

\begin{figure}[t]
\centering
\begin{tabular}{cc}
\includegraphics[width=7.7cm]{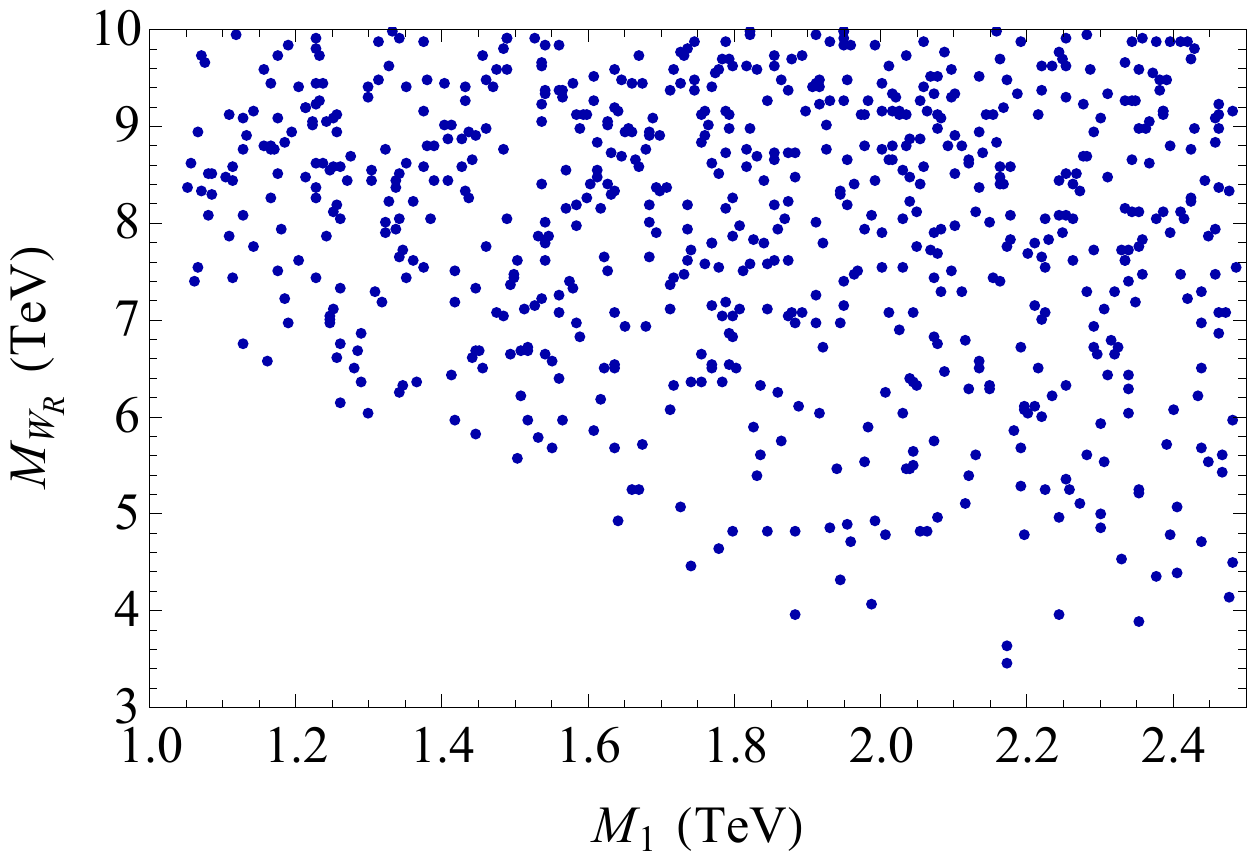} \qquad \qquad &
\includegraphics[width=7.7cm]{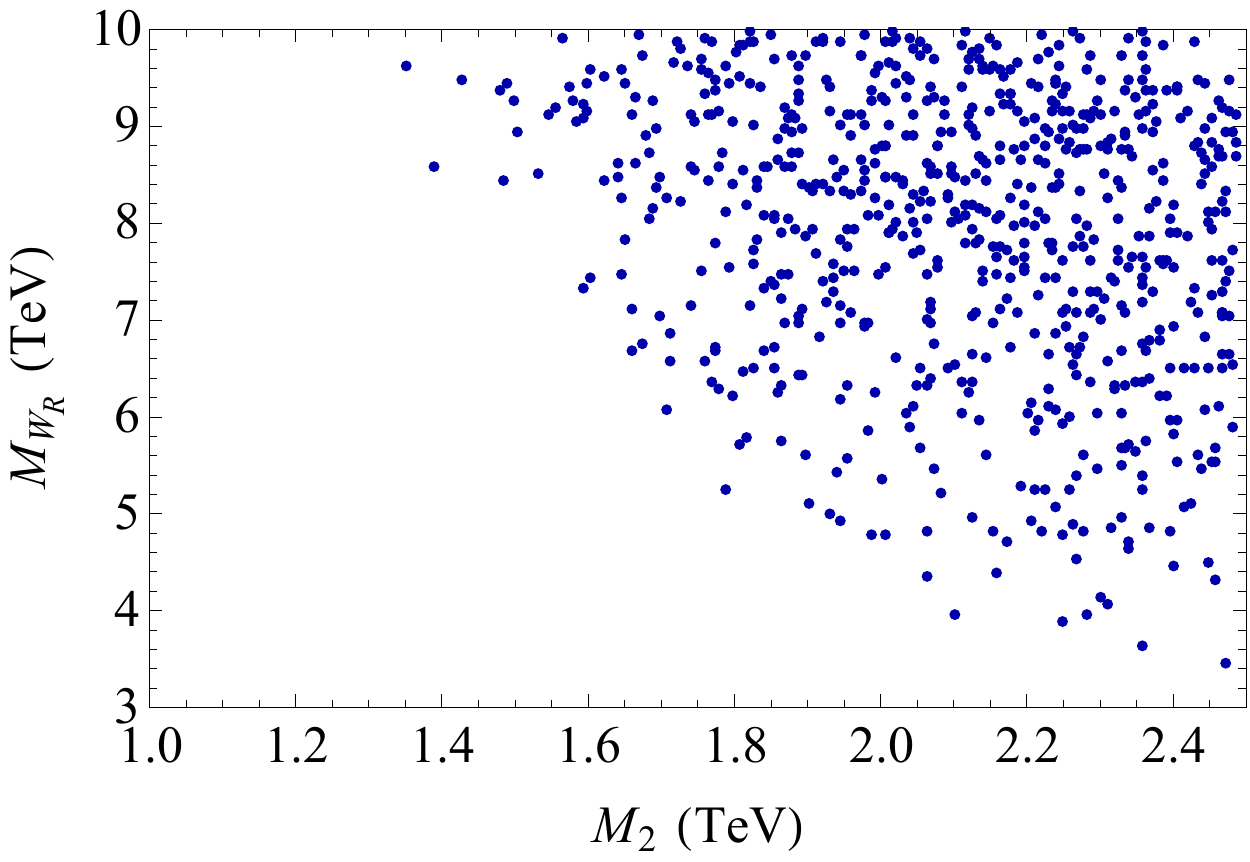}\\
(a) & (b) \\
\includegraphics[width=7.7cm]{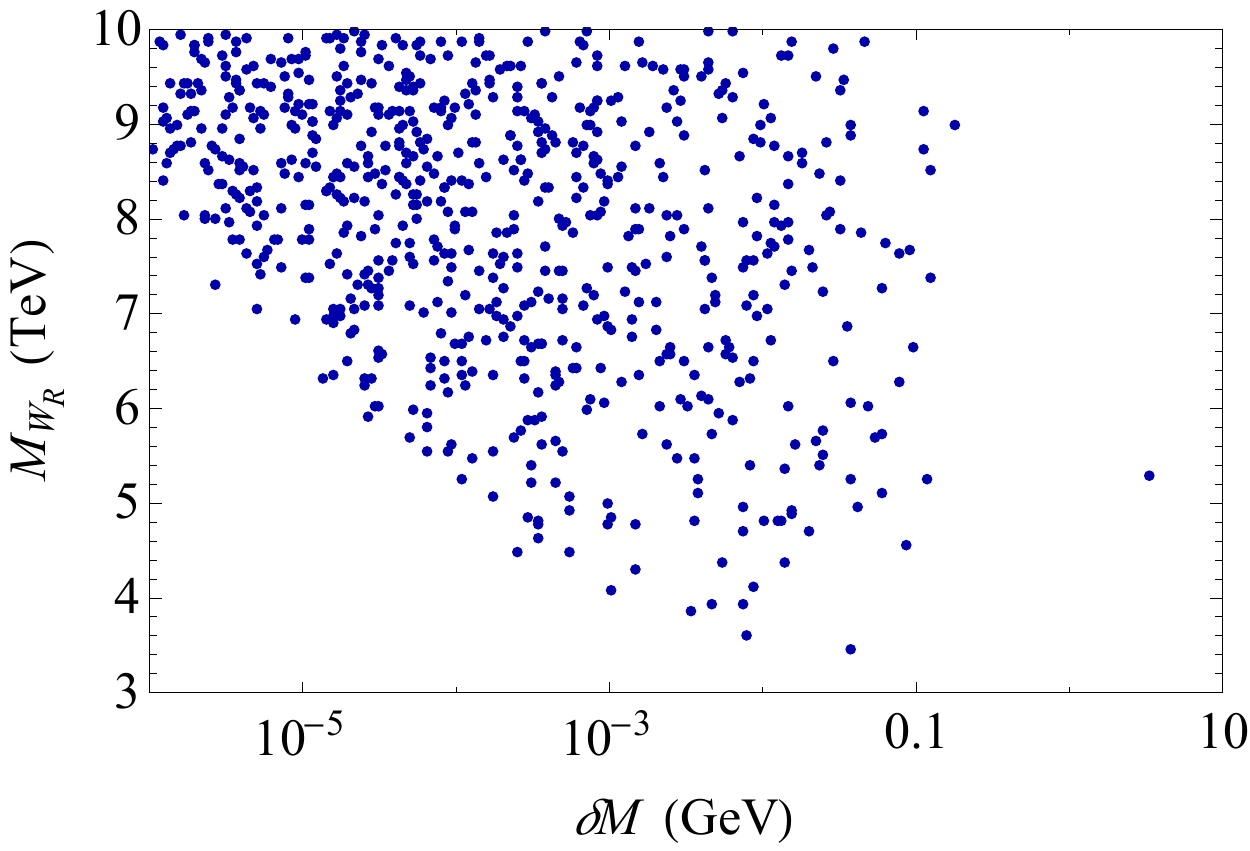} \qquad \qquad &
\includegraphics[width=7.7cm]{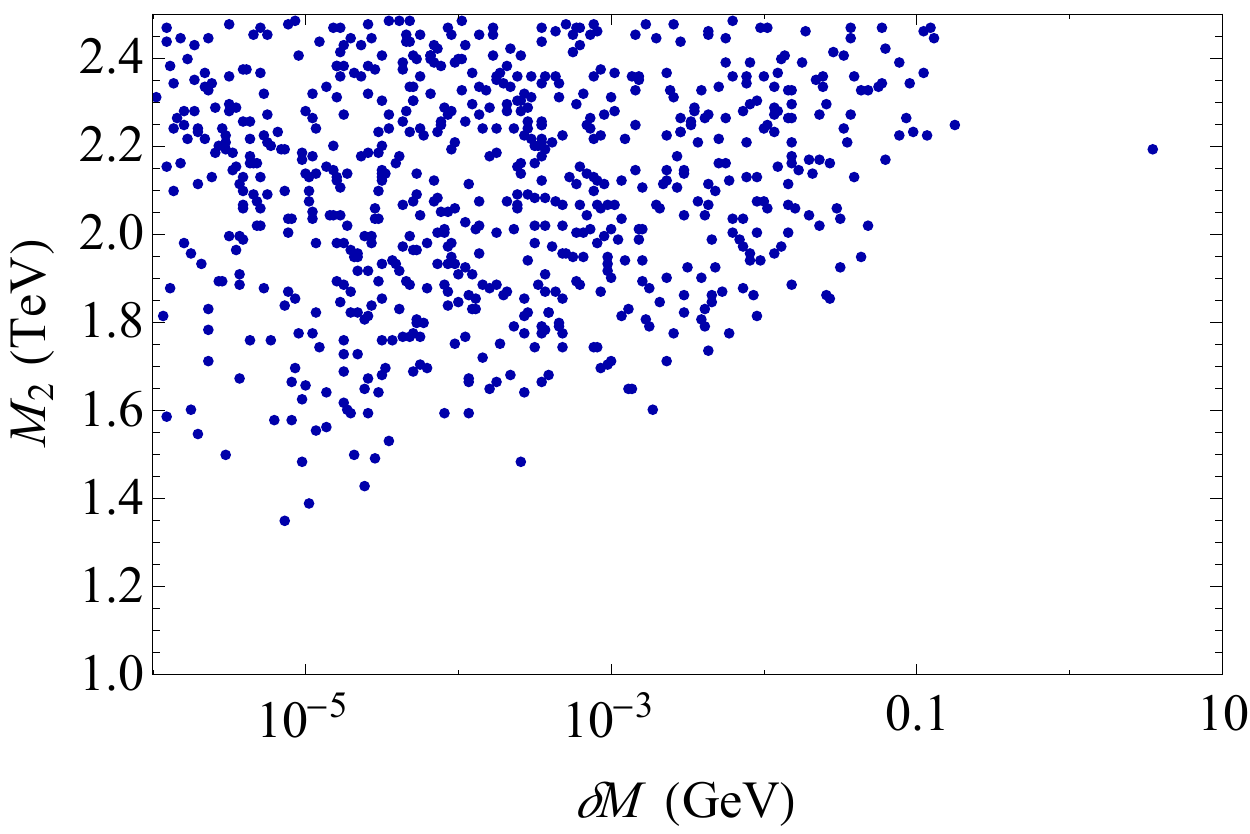}\\
(c) & (d) \\
\includegraphics[width=7.7cm]{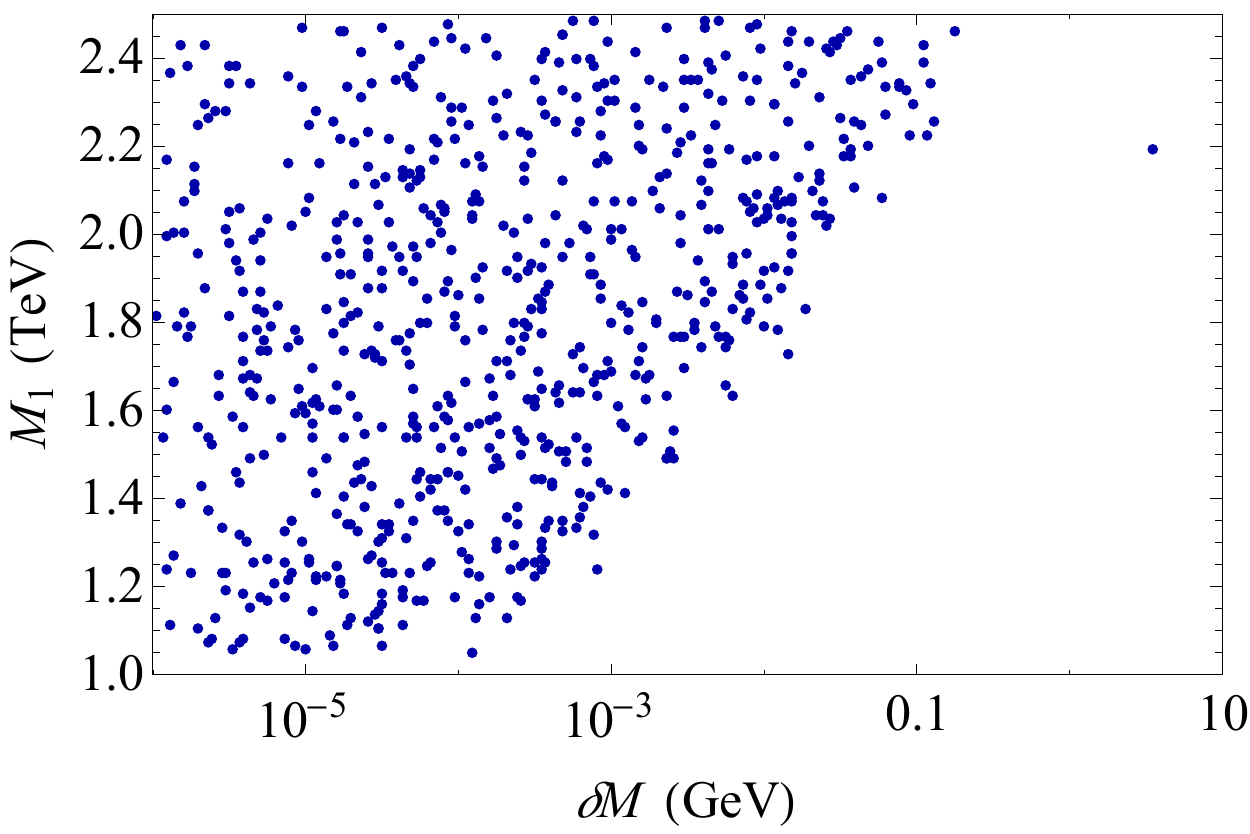} \qquad \qquad &
\includegraphics[width=7.7cm]{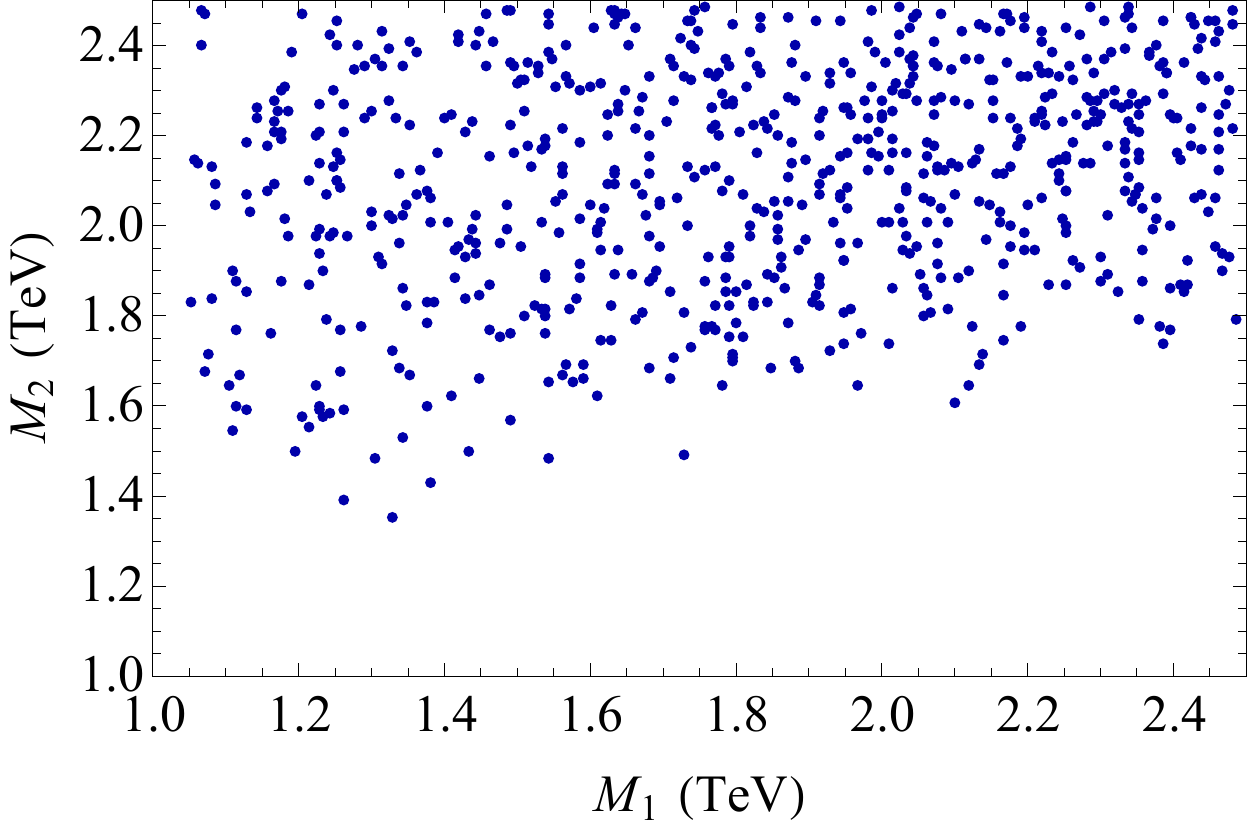} \\
(e) & (f) \\
\end{tabular}
\caption{The allowed parameter space in our TeV-scale L-R seesaw model, satisfying all the constraints in the lepton sector, including the neutrino oscillation data, LFV and leptogenesis constraints.}\label{fig:Y3}
\end{figure}

The allowed model parameter space yielding the correct value of lepton asymmetry in Figure~\ref{fig:Y2} is {\it not} just one exceptional fine-tuned situation. Instead, there exist several realistic numerical fits satisfying all the experimental constraints as well as yielding successful leptogenesis in our TeV-scale L-R model. This is demonstrated in Figures~\ref{fig:Y3} (a)-(f), where we show the allowed parameter space in terms of correlations between various relevant model parameters, namely $M_{W_R}$ and the Majorana mass parameters $\delta M$, $M_1$ and $M_2$ [cf.~Eq.~(\ref{eq:texture})]. It is clear that we have several solutions
with $M_{W_R}$ well below the previous lower bound of 18 TeV~\cite{hambye}.

\begin{figure}[t]
\centering
\includegraphics[width=15cm]{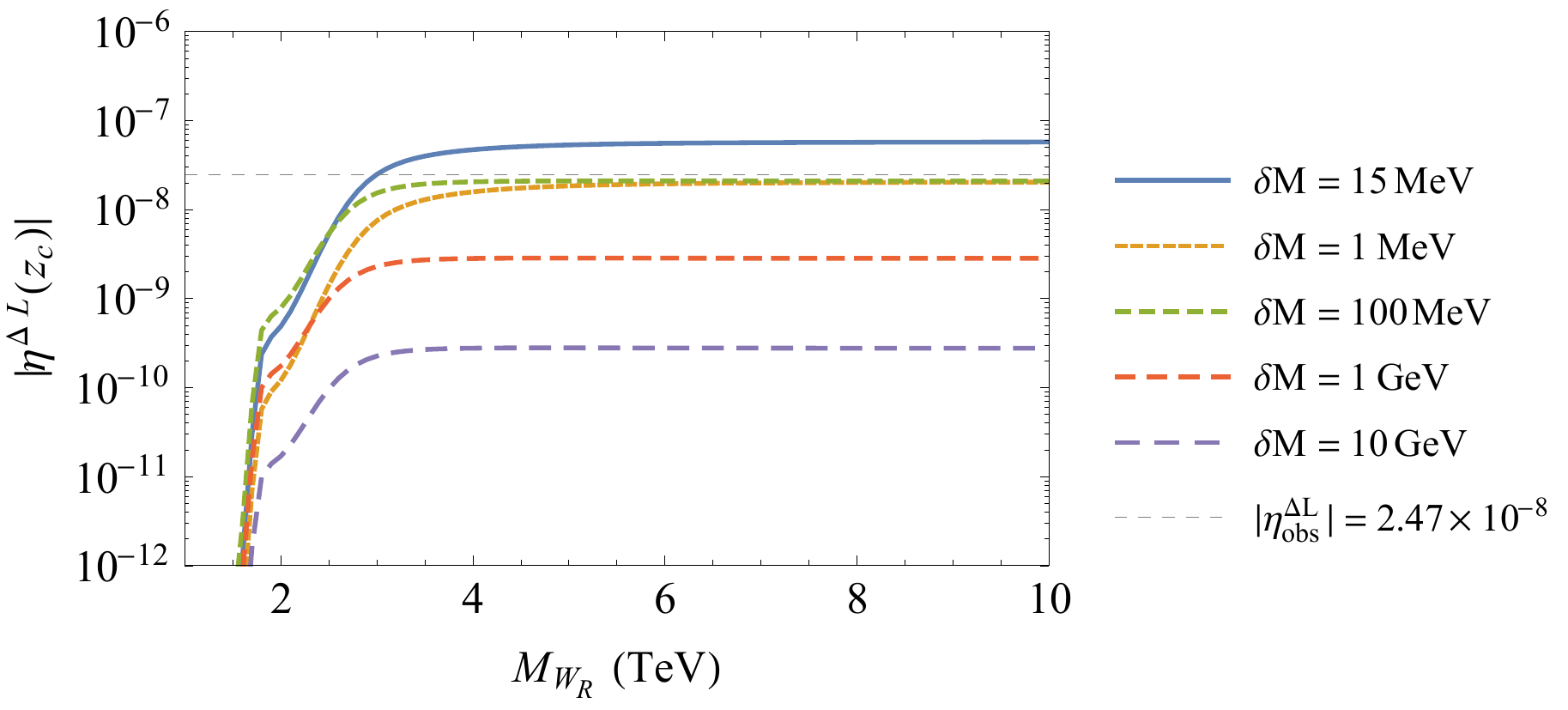}
\caption{The final lepton asymmetry as a function of the $W_R$ mass for various values of the degeneracy parameter. }\label{fig:WR}
\end{figure}

Finally, we examine if a weaker lower limit on the mass of $W_R$ can be set from leptogenesis constraints in the class of L-R seesaw models we are considering here.  The results are shown in Figure~\ref{fig:WR} for various choices of the degeneracy parameter $\delta M$, while the Yukawa couplings are kept fixed by choosing the neutrino fit of Section~\ref{sec:2a}. We find that the lower limit on $M_{W_R}$ is obtained when $\delta M$ is of the same order as the total decay width of the heavy neutrinos, i.e. when the resonant condition~\cite{Pilaftsis:1997dr} is exactly satisfied to yield the maximum $\CP$ asymmetry for a given Yukawa structure. For $\delta M\ll \Gamma_N$, the $\CP$-asymmetry gets suppressed  by $\delta M$ and vanishes as we approach the exact degeneracy limit $\delta M\to 0$, unless the contributions from the third heavy neutrino decays are significant. Similarly, for $\delta M\gg \Gamma_N$, we are in the hierarchical limit, when the $\CP$-asymmetry is dominantly produced from the decay of the lightest heavy neutrino, with no resonant enhancement effects.  From Figure~\ref{fig:WR}, we obtain a lower limit of $M_{W_R}>3$ TeV.

For the extreme case with all the three RH neutrino masses quasi-degenerate with each other and also with $W_R$, we obtain the absolute lower limit of $M_{W_R}>2.2$~TeV for successful leptogenesis. However, most of the allowed model parameter space is obtained for $M_{W_R}\gtrsim 3$ TeV or so, as shown in Figure~\ref{fig:Y3}.
\section{Discussions}\label{sec:4}
\begin{itemize}
\item {\bf Flavor Off-Diagonal Effects}: The Boltzmann equations (\ref{be1}) and (\ref{be2})
used in our leptogenesis analysis include only the diagonal flavor effects, i.e. the heavy neutrino and charged-lepton number densities were assumed to be diagonal. In general, flavor effects in resonant leptogenesis can give rise to three distinct physical phenomena~\cite{DMPT}: (i) resonant mixing between different heavy neutrino flavors, (ii) coherent oscillations between different heavy neutrino flavors, and (iii) coherences in the charged lepton sector. We have accounted for the first effect by using the effective Yukawa couplings, following the prescription given in~\cite{Pilaftsis:2003gt}. The third effect, namely charged-lepton coherences, is also partially taken into account in L-R models due to the inherent correlation between the charged lepton and neutrino Yukawa couplings. However, the flavor-diagonal Boltzmann equations (\ref{be1}) and (\ref{be2}) do not take into account the coherent oscillations between the heavy neutrino flavors, which could give rise to an additional source of $\CP$-asymmetry~\cite{ARS}.  It has been recently pointed out that the heavy neutrino oscillations around $z=1$ could enhance the final lepton asymmetry by up to a factor of two~\cite{DMPT}. In this sense, the lepton asymmetry results and the lower bound on $M_{W_R}$ presented in the previous section can be treated as conservative estimates, and could be further improved by generalizing the semi-classical Boltzmann equations (\ref{be1}) and (\ref{be2}) to the so-called `density matrix formalism'~\cite{sigl}\footnote{For an application of the `density matrix formalism' to study flavor effects in leptogenesis, see~\cite{flavor, DMPT}.} and by taking into account all the flavor effects in a consistent manner by means of a fully
flavor-covariant treatment~\cite{DMPT}.


\item {\bf Collider Prospects}: L-R models offer a unique probe of the Majorana nature of neutrinos via the smoking gun signature of same-sign dilepton plus two jets mediated by purely RH gauge currents~\cite{KS}. It must be emphasized here that, in the new class of models we have considered here, the quasi-degeneracy between the heavy neutrinos implies that, in the absence of the RH currents, the amplitude of the lepton number violating signal at colliders will be proportional to $M_DM_N^{-1}M_D^{\sf T}M_N^{-1}$, and hence, negligible, unless there is a resonant enhancement effect~\cite{bray}. On the other hand, in presence of the RH currents, there is a new contribution induced by the mixing between light and heavy neutrinos~\cite{CDM}, whose amplitude is only proportional to $M_DM_N^{-1}$. Hence, for relatively large Yukawa couplings as required for successful leptogenesis at low seesaw scale, the new `smoking gun' collider signal will be the so-called RL channel~\cite{CDM} with different helicity structure from the usual RR mode~\cite{KS}, and can be distinguished using simple kinematic variables~\cite{CDM, han}. Also, the observation of lepton number violating signal at the LHC would rule out high-scale models of leptogenesis~\cite{harz}, and might be able to shed light on the resonant leptogenesis scenarios discussed above.

\item {\bf CMS Excess}: While our paper was being finalized, the CMS collaboration announced their results for the $W_R$ search at $\sqrt s=8$ TeV LHC with $19.7~{\rm fb}^{-1}$ data~\cite{cms}.  Although they have claimed no significant excess of events over the SM expectation, and have presented 95\% CL exclusion contours in the $(M_{W_R},M_N)$ plane which now extends up to $M_{W_R}=3$ TeV,  there seems to be a $2.8\sigma$ excess in the invariant mass of the $eejj$ system around 2.1 TeV. This can be  interpreted as due to a heavy neutrino production and decay in LRSM with $M_{W_R}$ around 2.1 TeV, but with $g_R/g_L \simeq 0.6$ and $V_{eN}\simeq 0.9$~\cite{dep}. If the excess turns out to be statistically significant in future with more data (and independent scrutiny from ATLAS), it might be an evidence for L-R symmetry with high-scale parity breaking~\cite{CMP}. In this case, the leptogenesis constraints derived in this paper, assuming $g_R=g_L$, will be significantly weakened due to $g_R<g_L$. In fact, all the scattering and washout processes due to RH currents will be suppressed by a factor of $(g_L/g_R)^4\simeq 8$. This would mean that $M_{W_R}\simeq 2.1$ TeV is still compatible with the leptogenesis constraints derived here as well as other low-energy constraints given in Table~\ref{tab4}, especially from neutrinoless double beta decay.  A detailed phenomenological analysis of this $D$-parity breaking scenario in our class of L-R seesaw models and their correlation with successful leptogenesis will be presented in a future communication.
\end{itemize}
\section{Conclusion}\label{sec:5}
We have analyzed the leptogenesis constraints on the mass of the right-handed gauge boson in TeV-scale Left-Right Symmetric Models. While the existing bound of $M_{W_R}>18$ TeV applies for generic LRSM scenarios with small Yukawa couplings, we have found a significantly weaker bound of  $M_{W_R}>3$ TeV in  a new class of L-R seesaw models with relatively larger Yukawa couplings, which is consistent with charged-lepton and neutrino oscillation data. The key factors responsible for our result are: (i) specific textures of the Dirac and Majorana mass matrices to ensure that the neutrino and charged-lepton data are satisfied, even for large Yukawa couplings, (ii) suppressed dilution effect from $W_R$-mediated scatterings and decays, and (iii) inclusion of flavor effects in the lepton asymmetry calculation. This lower bound of $M_{W_R}>3$ TeV is obtained for the case $g_L=g_R$ and will be proportionately weaker for the $g_R < g_L$ case. The bounds obtained here are comparable to the existing collider and low energy bounds, and more importantly, makes it difficult to falsify leptogenesis at the LHC by the discovery of the $W_R$ alone. Determining other parameters of the theory such as masses of the right-handed neutrinos and the Yukawa couplings could however help us to test the validity of the leptogenesis hypothesis.
\section*{Acknowledgment} We gratefully acknowledge helpful discussions with Thomas Hambye, Apostolos Pilaftsis and Daniele Teresi. We also thank Un-ki Yang for clarifying remarks on the CMS $W_R$ search. The work of P. S. B. D. is supported by
 the Lancaster-Manchester-Sheffield Consortium for Fundamental
 Physics under STFC grant ST/J000418/1. The works of C.-H. L. and R. N. M. are supported by the U.S. National Science Foundation grant No. PHY-1315155.

\appendix
\section{Reduced Cross Sections}\label{app:A}
The reduced cross sections for the dominant processes involving RH currents which determine the scattering rates $\gamma^{S_R}_{l\alpha}$ in Eqs.~(\ref{be1}) and (\ref{be2}) are given by~\cite{hambye, Blanchet:2010kw}
\begin{eqnarray}
\hat{\sigma}^{Nl_R}_{\bar{u}_Rd_R} (x) & = & \frac{9g_R^4}{48\pi x}\frac{1-3x^2+2x^3}{[(x-a_R)^2+a_Rc_R]} \, , \nonumber \\
\hat{\sigma}^{N\bar{u}_R}_{e_R\bar{d}_R} (x) & = & \frac{9g_R^4}{8\pi x}\int_{1-x}^0 du \frac{(x+u)(x+u-1)}{(u-a_R)^2} \, , \nonumber \\
\hat{\sigma}^{Nd_R}_{e_R u_R}(x) & =& \frac{9g_R^4}{8\pi a_R}\frac{(1-x)^2}{(x+a_R-1)} \, ,
\end{eqnarray}
where $a_R=(M_{W_R}/m_{N_1})^2,~c_R=(\Gamma_{W_R}/m_{N_1})^2$.

There are analogous $\Delta L=1$ processes mediated by the scalar fields.\footnote{Above the electroweak phase transition scale, the SM-like Higgs doublet does not have a VEV, and hence, there is no mixing between LH and RH neutrino fields induced in the gauge interactions of the theory.  Thus, the only relevant SM contribution comes from the Yukawa interactions.} The reduced cross sections for these processes are given by~\cite{Pilaftsis:2003gt, Luty:1992un}
\begin{eqnarray}
\hat{\sigma}^{N_\alpha L_l}_{Qu^c}(x) &=& 3\alpha_u\left(\widehat{\maf{h}}^*_{l\alpha}\widehat{\maf{h}}_{l\alpha}+\widehat{\maf{h}}^{c*}_{l\alpha}\widehat{\maf{h}}^c_{l\alpha}\right)\left(1-\frac{a_\alpha}{x}\right)^2 \, ,\nonumber \\
\hat{\sigma}^{N_\alpha u^c}_{LQ^c}(x) &=& \hat{\sigma}^{N_\alpha Q}_{Lu}(x) = 3\alpha_u\left(\widehat{\maf{h}}^*_{l\alpha}\widehat{\maf{h}}_{l\alpha}+\widehat{\maf{h}}^{c*}_{l\alpha}\widehat{\maf{h}}^c_{l\alpha}\right)\left(1+\frac{a_\alpha}{x}\left[\ln\left(1+\frac{x-a_\alpha}{a_r}\right)-1\right]\right) \, ,
\end{eqnarray}
where $\alpha_u={\rm Tr}(h_u^\dag h_u)/4\pi\simeq \alpha_Wm_t^2/2M_W^2$, $a_\alpha=(m_{N_\alpha}/m_{N_1})^2\simeq 1$ for quasi-degenerate heavy neutrinos, and $a_r=(m_{\rm IR}/m_{N_1})^2$, where $m_{\rm IR}$ is an IR regulator for the $t$-channel processes involving massless particles. Here, $m_{\rm IR}$ is chosen to be the Higgs thermal mass $M_\phi(T) = [2D(T^2 - T_c^2)]^{1/2}$ for $T>T_c$,
where $T_c$ is the critical temperature given by Eq.~(\ref{tc}), and for $T\leq T_c$, $m_{\rm IR}=M_H$ (zero-temperature Higgs mass).

There are additional $\Delta L=1$ reactions involving the SM gauge bosons $V_\mu=B_\mu, W_\mu^a$ in the initial or final states. To leading order in $a_r$, the corresponding $\CP$-conserving reduced cross sections are given by~\cite{Pilaftsis:2003gt}
\begin{eqnarray}
\hat{\sigma}^{N_\alpha V_\mu}_{L_l\phi} & = & \frac{n_V g_V^2}{8\pi x}\left(\widehat{\maf{h}}^*_{l\alpha}\widehat{\maf{h}}_{l\alpha}+\widehat{\maf{h}}^{c*}_{l\alpha}\widehat{\maf{h}}^c_{l\alpha}\right)\left[\frac{(x+a_\alpha)^2}{x-a_\alpha+2a_r}\ln\left(1+\frac{x-a_\alpha}{a_r}\right)\right] \, , \nonumber \\
\hat{\sigma}^{N_\alpha L_l}_{\phi^\dag V_\mu} & = & \frac{n_V g_V^2}{16\pi x^2}\left(\widehat{\maf{h}}^*_{l\alpha}\widehat{\maf{h}}_{l\alpha}+\widehat{\maf{h}}^{c*}_{l\alpha}\widehat{\maf{h}}^c_{l\alpha}\right)\left[(5x-a_\alpha)(a_\alpha-x)+2(x^2+xa_\alpha-a_\alpha^2)\ln\left(1+\frac{x-a_\alpha}{a_r}\right)\right] \, ,\nonumber \\
\hat{\sigma}^{N_\alpha \phi^\dag}_{L_l V_\mu} & = & \frac{n_V g_V^2}{16\pi x^2}\left(\widehat{\maf{h}}^*_{l\alpha}\widehat{\maf{h}}_{l\alpha}+\widehat{\maf{h}}^{c*}_{l\alpha}\widehat{\maf{h}}^c_{l\alpha}\right))(x-a_\alpha)\left[x-3a_\alpha+4a_\alpha\ln\left(1+\frac{x-a_\alpha}{a_r}\right)\right] \, ,
\end{eqnarray}
where $g_V=g',g_L$ and $n_V=1,3$ for $V_\mu=B_\mu, W_\mu^a$, respectively. It turns out that these $\Delta L=1$ scatterings involving only the SM fields are sub-dominant compared to those involving $W_R$ in our L-R seesaw model (cf.~Figure~\ref{fig:rate}).




\begin{thebibliography}{99}


\bibitem{seesaw} P. Minkowski, Phys. Lett. B {\bf 67}, 421 (1977);
R. N. Mohapatra and G. Senjanovi\'{c}, Phys. Rev. Lett. {\bf 44}, 912 (1980);
T. Yanagida, {\it Workshop on unified theories and baryon number in the universe}, edited by A. Sawada and A. Sugamoto, KEK, Tsukuba (1979);
M. Gell-Mann, P. Ramond and R. Slansky, {\it Supergravity}, edited by P. Van Niewenhuizen and D. Freedman, North Holland, Amsterdam (1980).



\bibitem{LR} J.C. Pati and A. Salam, Phys. Rev. D {\bf 10}, 275 (1974);
R. N. Mohapatra and J. C. Pati, Phys. Rev. D {\bf 11}, 566 (1975); Phys. Rev. D {\bf 11},  2558 (1975);
G. Senjanovi\'{c} and R. N. Mohapatra, Phys. Rev. D {\bf 12} 1502 (1975).

\bibitem{so10} H. Georgi, {\it Particles and Fields}, edited by by C. Carlson, AIP (1975);
H. Friztsch and P. Minkowski, Ann. Phys. {\bf 93}, 193 (1975).

 \bibitem{ours}  P.~S.~B.~Dev, C.-H.~Lee and R.~N.~Mohapatra,
  Phys.\ Rev.\ D {\bf 88}, 093010 (2013) [arXiv:1309.0774 [hep-ph]].

\bibitem{lepto} M.~Fukugita and T.~Yanagida,
  Phys.\ Lett.\ B {\bf 174}, 45 (1986).

\bibitem{reviews} S.~Davidson, E.~Nardi and Y.~Nir,
  Phys.\ Rept.\  {\bf 466}, 105 (2008)
  [arXiv:0802.2962 [hep-ph]];
G.~C.~Branco, R.~G.~Felipe and F.~R.~Joaquim,
  Rev.\ Mod.\ Phys.\  {\bf 84}, 515 (2012)
  [arXiv:1111.5332 [hep-ph]]; 
  S.~Blanchet and P.~Di Bari,
  New J.\ Phys.\  {\bf 14}, 125012 (2012)
  [arXiv:1211.0512 [hep-ph]]; 
C.~S.~Fong, E.~Nardi and A.~Riotto,
  Adv.\ High Energy Phys.\  {\bf 2012}, 158303 (2012)
  [arXiv:1301.3062 [hep-ph]]. 
 
 

\bibitem{sakharov} A.~D.~Sakharov,
 JETP Lett.\  {\bf 5}, 24 (1967).

\bibitem{Kuzmin:1985mm}
  V.~A.~Kuzmin, V.~A.~Rubakov and M.~E.~Shaposhnikov,
  Phys.\ Lett.\ B {\bf 155}, 36 (1985).

\bibitem{Pilaftsis:1997dr}
  A.~Pilaftsis,
  Nucl.\ Phys.\ B {\bf 504}, 61 (1997)
  [hep-ph/9702393];
  A.~Pilaftsis,
  Phys.\ Rev.\ D {\bf 56}, 5431 (1997)
  [hep-ph/9707235].

\bibitem{Pilaftsis:2003gt}
A.~Pilaftsis and T.~E.~J.~Underwood,
  Nucl.\ Phys.\ B {\bf 692}, 303 (2004)
  [hep-ph/0309342].

\bibitem{Liu:1993tg}
  J.~Liu and G.~Segr\`{e},
  Phys.\ Rev.\ D {\bf 48}, 4609 (1993)
  [hep-ph/9304241].

\bibitem{Flanz:1994yx}
  M.~Flanz, E.~A.~Paschos and U.~Sarkar,
  Phys.\ Lett.\ B {\bf 345}, 248 (1995)
  [Erratum-ibid.\ B {\bf 382}, 447 (1996)]
  [hep-ph/9411366];
  L.~Covi, E.~Roulet and F.~Vissani,
  Phys.\ Lett.\ B {\bf 384}, 169 (1996)
  [hep-ph/9605319].

\bibitem{Carlier:1999ac}
  S.~Carlier, J.-M.~Frere and F.~S.~Ling,
  Phys.\ Rev.\ D {\bf 60}, 096003 (1999)
  [hep-ph/9903300].

 \bibitem{hambye} J.-M.~Frere, T.~Hambye and G.~Vertongen,
  JHEP {\bf 0901}, 051 (2009)
  [arXiv:0806.0841 [hep-ph]].

\bibitem{pdg}
  J. Beringer {\it et al.} (Particle Data Group),
  Phys. Rev. D {\bf 86}, 010001 (2012)
  [{\tt http://pdg.lbl.gov/}].




\bibitem{DMPT} P.~S.~B.~Dev, P.~Millington, A.~Pilaftsis and D.~Teresi,
  Nucl.\ Phys.\ B {\bf 886}, 569 (2014)
  [arXiv:1404.1003 [hep-ph]];  {\em ibid.} arXiv:1409.8263 [hep-ph]. 

\bibitem{charge} R. N. Mohapatra and R. E. Marshak, Phys. Lett. {\bf B 91}, 222 (1980); 
A. Davidson, Phys. Rev. D {\bf 20}, 776 (1979). 


 \bibitem{CMP} D.~Chang, R.~N.~Mohapatra and M.~K.~Parida,
  Phys.\ Rev.\ Lett.\  {\bf 52}, 1072 (1984).

\bibitem{type2} 
J.~Schechter and J.~W.~F.~Valle,
  Phys.\ Rev.\ D {\bf 22}, 2227 (1980);
T.~P.~Cheng and L.-F.~Li, Phys. Rev. {\bf D 22}, 2860 (1980);
G.~Lazarides, Q.~Shafi and C.~Wetterich, Nucl. Phys. {\bf B 181}, 287 (1981);
R.~N.~Mohapatra and G.~Senjanovi\'{c}, Phys. Rev. {\bf D 23}, 165 (1981).

\bibitem{type2lepto} R.~Gonzalez Felipe, F.~R.~Joaquim and H.~Serodio,
  Int.\ J.\ Mod.\ Phys.\ A {\bf 28}, 1350165 (2013) [arXiv:1301.0288 [hep-ph]]; 
D.~Aristizabal Sierra, M.~Dhen and T.~Hambye,
  JCAP {\bf 1408}, 003 (2014)
  [arXiv:1401.4347 [hep-ph]].
  
\bibitem{RGE} R.~Gonzalez Felipe, F.~R.~Joaquim and B.~M.~Nobre,
  Phys.\ Rev.\ D {\bf 70}, 085009 (2004) [hep-ph/0311029]; 
G.~C.~Branco, R.~Gonzalez Felipe, F.~R.~Joaquim and B.~M.~Nobre,
  Phys.\ Lett.\ B {\bf 633}, 336 (2006) [hep-ph/0507092].
  
 \bibitem{Pilaftsis:2005rv}
  A.~Pilaftsis and T.~E.~J.~Underwood,
  Phys.\ Rev.\ D {\bf 72}, 113001 (2005)
  [hep-ph/0506107].

\bibitem{Deppisch:2010fr}
F.~F.~Deppisch and A.~Pilaftsis,
  Phys.\ Rev.\ D {\bf 83}, 076007 (2011)
  [arXiv:1012.1834 [hep-ph]].
 

\bibitem{Forero:2014bxa}
  D.~V.~Forero, M.~Tortola and J.~W.~F.~Valle,
  arXiv:1405.7540 [hep-ph].

\bibitem{fcnc} R.~N.~Mohapatra, G.~Senjanovi\'{c} and M.~D.~Tran, Phys. Rev. D {\bf 28},
546 (1983);
G.~Ecker, W.~Grimus and H.~Neufeld, Phys. Lett. B {\bf 127}, 365 (1983) [Erratum-ibid. B {\bf 132}, 467 (1983)];
Y.~Zhang, H.~An, X.~Ji and R.~N.~Mohapatra,
  Nucl.\ Phys.\ B {\bf 802}, 247 (2008)
  [arXiv:0712.4218 [hep-ph]];
A.~Maiezza, M.~Nemevsek, F.~Nesti and G.~Senjanovic,
  Phys.\ Rev.\ D {\bf 82}, 055022 (2010)
  [arXiv:1005.5160 [hep-ph]].

\bibitem{pal} P.~B.~Pal,
  Nucl.\ Phys.\ B {\bf 227}, 237 (1983).

 \bibitem{rnm-92} R.~N.~Mohapatra,
  Phys.\ Rev.\ D {\bf 46}, 2990 (1992).

\bibitem{mueg-L} W.~J.~Marciano and A.~I.~Sanda,
  Phys.\ Lett.\ B {\bf 67}, 303 (1977);
T.~P.~Cheng and L.-F.~Li,
  Phys.\ Rev.\ Lett.\  {\bf 45}, 1908 (1980).

\bibitem{mueg-R} Riazuddin, R.~E.~Marshak and R.~N.~Mohapatra,
  Phys.\ Rev.\ D {\bf 24}, 1310 (1981).

\bibitem{Adam:2013mnn}
  J.~Adam {\it et al.}  [MEG Collaboration],
  Phys.\ Rev.\ Lett.\  {\bf 110}, 201801 (2013)
  [arXiv:1303.0754 [hep-ex]].

\bibitem{alonso} R.~Alonso, M.~Dhen, M.~B.~Gavela and T.~Hambye,
  JHEP {\bf 1301}, 118 (2013)
  [arXiv:1209.2679 [hep-ph]].

\bibitem{mue-others} D.~N.~Dinh, A.~Ibarra, E.~Molinaro and S.~T.~Petcov,
  JHEP {\bf 1208}, 125 (2012)
  [Erratum-ibid.\  {\bf 1309}, 023 (2013)]
  [arXiv:1205.4671 [hep-ph]];
A.~Ilakovac, A.~Pilaftsis and L.~Popov,
  Phys.\ Rev.\ D {\bf 87}, 053014 (2013)
  [arXiv:1212.5939 [hep-ph]];
 A.~Abada, M.~E.~Krauss, W.~Porod, F.~Staub, A.~Vicente and C.~Weiland,
  arXiv:1408.0138 [hep-ph].

\bibitem{mueTi}
  J.~Kaulard {\it et al.}  [SINDRUM II Collaboration],
  Phys.\ Lett.\ B {\bf 422}, 334 (1998);
  P. Wintz,
  in {\it Proceedings of the First International Symposium on
    Lepton and Baryon Number Violation},
  eds. H. V. Klapdor-Kleingrothaus and I. V. Krivosheina,
  Institute of Physics Publishing, Bristol, p. 534 (1998).

\bibitem{mueAu}
  W.~H.~Bertl {\it et al.}  [SINDRUM II Collaboration],
  Eur.\ Phys.\ J.\ C {\bf 47}, 337 (2006).

\bibitem{muePb}
  W.~Honecker {\it et al.}  [SINDRUM II Collaboration],
  Phys.\ Rev.\ Lett.\  {\bf 76}, 200 (1996).


\bibitem{LR-mixed} M.~Hirsch, H.~V.~Klapdor-Kleingrothaus and O.~Panella,
  Phys.\ Lett.\ B {\bf 374}, 7 (1996)
  [hep-ph/9602306];
J. Barry and W. Rodejohann, JHEP {\bf 1309}, 153 (2013) [arXiv:1303.6324 [hep-ph]];
W.-C.~Huang and J.~Lopez-Pavon,
  Eur.\ Phys.\ J.\ C {\bf 74}, 2853 (2014)
  [arXiv:1310.0265 [hep-ph]];
P.~S.~B.~Dev, S.~Goswami and M.~Mitra,
  arXiv:1405.1399 [hep-ph].

\bibitem{racah} G. Racah, Nuovo Cimento {\bf 14}, 322 (1937);
W. H. Furry, Phys. Rev. {\bf 56}, 1184 (1939).

\bibitem{RH} R. N. Mohapatra and G. Senjanovi\'{c}, in Ref.~\cite{seesaw};
Phys. Rev. D {\bf 23}, 165 (1981);
R.~N.~Mohapatra and J.~D.~Vergados,
  Phys.\ Rev.\ Lett.\  {\bf 47}, 1713 (1981);
C.~E.~Picciotto and M.~S.~Zahir,
  Phys.\ Rev.\ D {\bf 26}, 2320 (1982);
 V.~Tello, M.~Nemevsek, F.~Nesti, G.~Senjanovic and F.~Vissani,
  Phys.\ Rev.\ Lett.\  {\bf 106}, 151801 (2011)
  [arXiv:1011.3522 [hep-ph]].

\bibitem{NME} G.~Pantis, F.~Simkovic, J.~D.~Vergados and A.~Faessler,
  Phys.\ Rev.\ C {\bf 53}, 695 (1996)
  [nucl-th/9612036]; J.~Suhonen and O.~Civitarese,
  Phys.\ Rept.\  {\bf 300}, 123 (1998);
A.~Meroni, S.~T.~Petcov and F.~Simkovic,
  JHEP {\bf 1302}, 025 (2013)
  [arXiv:1212.1331];
P.~S.~B. Dev, S.~Goswami, M.~Mitra and W.~Rodejohann,
  Phys.\ Rev.\ D {\bf 88}, 091301 (2013) [arXiv:1305.0056 [hep-ph]].

\bibitem{gerda}
  M.~Agostini {\it et al.}
  [GERDA Collaboration],
  Phys.\ Rev.\ Lett.\  {\bf 111}, 122503 (2013)
  [arXiv:1307.4720 [nucl-ex]].

\bibitem{kamland} A.~Gando {\it et al.}  [KamLAND-Zen Collaboration],
  Phys.\ Rev.\ Lett.\  {\bf 110}, 062502 (2013)
  [arXiv:1211.3863 [hep-ex]];
I. Shimizu, Talk at Neutrino 2014, Boston, USA (June 6, 2014).

\bibitem{cuo}  E.~Andreotti {\it et al.} [CUORICINO Collaboration],
  Astropart.\ Phys.\  {\bf 34}, 822 (2011)
  [arXiv:1012.3266 [nucl-ex]].

\bibitem{meg2} A.~M.~Baldini {\it et al.} [MEG Collaboration],
  arXiv:1301.7225 [physics.ins-det].


\bibitem{gerda2}  K.~H.~Ackermann {\it et al.}  [GERDA Collaboration],
  Eur.\ Phys.\ J.\ C {\bf 73}, 2330 (2013)
  [arXiv:1212.4067 [physics.ins-det]].

\bibitem{majorana}  N.~Abgrall {\it et al.}  [Majorana Collaboration],
  Adv.\ High Energy Phys.\  {\bf 2014}, 365432 (2014)
  [arXiv:1308.1633 [physics.ins-det]].

\bibitem{exo}  M.~Auger {\it et al.} [EXO Collaboration],
  JINST {\bf 7}, P05010 (2012)
  [arXiv:1202.2192 [physics.ins-det]].

\bibitem{cuore}  D.~R.~Artusa {\it et al.}  [CUORE Collaboration],
  arXiv:1402.6072 [physics.ins-det].

 


\bibitem{Pilaftsis:2008qt}
  A.~Pilaftsis,
  Phys.\ Rev.\ D {\bf 78}, 013008 (2008)
  [arXiv:0805.1677 [hep-ph]].

\bibitem{Blanchet:2009bu}
  S.~Blanchet, Z.~Chacko, S.~S.~Granor and R.~N.~Mohapatra,
  Phys.\ Rev.\ D {\bf 82}, 076008 (2010)
  [arXiv:0904.2174 [hep-ph]];

\bibitem{Blanchet:2010kw}
  S.~Blanchet, P.~S.~B.~Dev and R.~N.~Mohapatra,
  Phys.\ Rev.\ D {\bf 82}, 115025 (2010)
  [arXiv:1010.1471 [hep-ph]].







\bibitem{Cline:1993bd}
  J.~M.~Cline, K.~Kainulainen and K.~A.~Olive,
  Phys.\ Rev.\ D {\bf 49}, 6394 (1994)
  [hep-ph/9401208].

\bibitem{Buchmuller:2004nz}
  W.~Buchmuller, P.~Di Bari and M.~Plumacher,
  Annals Phys.\  {\bf 315}, 305 (2005)
  [hep-ph/0401240].


\bibitem{planck}  P.~A.~R.~Ade {\it et al.}  [Planck Collaboration],
  arXiv:1303.5076 [astro-ph.CO].


\bibitem{ARS}
  E.~K.~Akhmedov, V.~A.~Rubakov and A.~Y.~Smirnov,
  Phys.\ Rev.\ Lett.\  {\bf 81}, 1359 (1998)
  [hep-ph/9803255].


\bibitem{sigl} G.~Sigl and G.~Raffelt,
  Nucl.\ Phys.\ B {\bf 406}, 423 (1993).

\bibitem{flavor}
  A.~Abada, S.~Davidson, F.-X.~Josse-Michaux, M.~Losada and A.~Riotto,
  JCAP {\bf 0604}, 004 (2006)
  [hep-ph/0601083];
  E.~Nardi, Y.~Nir, E.~Roulet and J.~Racker,
  JHEP {\bf 0601}, 164 (2006)
  [hep-ph/0601084];
  A.~Abada, S.~Davidson, A.~Ibarra, F.-X.~Josse-Michaux,
  M.~Losada and A.~Riotto,
  JHEP {\bf 0609}, 010 (2006)
  [hep-ph/0605281];
  S.~Blanchet and P.~Di Bari,
  JCAP {\bf 0703}, 018 (2007)
  [hep-ph/0607330];
  A.~De Simone and A.~Riotto,
  JCAP {\bf 0702}, 005 (2007)
  [hep-ph/0611357];
  S.~Blanchet, P.~Di Bari, D.~A.~Jones and L.~Marzola,
  JCAP {\bf 1301}, 041 (2013)
  [arXiv:1112.4528 [hep-ph]].


\bibitem{KS} W.-Y. Keung and G. Senjanov\'{i}c, Phys. Rev. Lett. {\bf 50}, 1427 (1983).



\bibitem{bray} S.~Bray, J.~S.~Lee and A.~Pilaftsis,
  Nucl.\ Phys.\ B {\bf 786}, 95 (2007)
  [hep-ph/0702294 [hep-ph]];
P.~S.~B.~Dev, A.~Pilaftsis and U.~K.~Yang,
  Phys.\ Rev.\ Lett.\  {\bf 112}, 081801 (2014)
  [arXiv:1308.2209 [hep-ph]].

\bibitem{CDM} C.-Y. Chen, P. S. B. Dev and R. N. Mohapatra, Phys. Rev. D 88, 033014 (2013) [arXiv:1306.2342 [hep-ph]];  P.~S.~B.~Dev and R.~N.~Mohapatra,
  arXiv:1308.2151 [hep-ph].

\bibitem{han} T.~Han, I.~Lewis, R.~Ruiz and Z.~-g.~Si,
  Phys.\ Rev.\ D {\bf 87}, 035011 (2013)
  [Erratum-ibid.\ D {\bf 87}, 039906 (2013)]
  [arXiv:1211.6447 [hep-ph]].


\bibitem{harz} F.~F.~Deppisch, J.~Harz and M.~Hirsch,
  Phys.\ Rev.\ Lett.\  {\bf 112}, 221601 (2014)
  [arXiv:1312.4447 [hep-ph]].

\bibitem{cms} V.~Khachatryan {\it et al.}  [CMS Collaboration],
  arXiv:1407.3683 [hep-ex].

\bibitem{dep} F.~F.~Deppisch, T.~E.~Gonzalo, S.~Patra, N.~Sahu and U.~Sarkar,
  Phys. Rev. D {\bf 90}, 053014 (2014) [arXiv:1407.5384 [hep-ph]];
M.~Heikinheimo, M.~Raidal and C.~Spethmann,
  arXiv:1407.6908 [hep-ph];
J.~A.~Aguilar-Saavedra and F.~R.~Joaquim, arXiv:1408.2456 [hep-ph].









\bibitem{Luty:1992un}
  M.~A.~Luty,
  Phys.\ Rev.\ D {\bf 45}, 455 (1992).


\end{thebibliography}
\end{document}